\documentclass[%
 aip,
 amsmath,amssymb,
 reprint,%
]{revtex4-1}

\usepackage{graphicx}
\usepackage{dcolumn}
\usepackage{bm}
\newcommand{\dd}{\mathrm{d}}
\usepackage{siunitx}

\usepackage[utf8]{inputenc}
\usepackage[T1]{fontenc}
\usepackage{mathptmx}
\usepackage{etoolbox}
\usepackage{xcolor}

\usepackage{graphicx}
\usepackage{amsmath}
\usepackage{color}
\usepackage{sidecap}
\usepackage{lineno}
\usepackage{cancel}

\makeatletter
\def\@email#1#2{%
 \endgroup
 \patchcmd{\titleblock@produce}
 {\frontmatter@RRAPformat}
 {\frontmatter@RRAPformat{\produce@RRAP{*#1\href{mailto:#2}{#2}}}\frontmatter@RRAPformat}
 {}{}
}%
\makeatother

\newcommand{\beq}{\begin{eqnarray}}
\newcommand{\eeq}{\end{eqnarray}}

\newcommand{\vp}{{v_\perp}}

\newcommand{\vv}{{\bf v}}

\newcommand{\scr}[1]{_{\mbox{\protect\scriptsize #1}} }
\newcommand{\sscr}[1]{^{\mbox{\protect\scriptsize #1}} }
\newcommand{\E}{{\bf E}}
\newcommand{\B}{{\bf B}}

\def\hblabel#1{\label{#1}}

\newcommand{\vz}{v_{\parallel}}
\newcommand{\fd}[2]{\frac{\displaystyle #1}{\displaystyle #2}}

\newcommand{\pz}{{p_{e\parallel}}}
\newcommand{\pp}{{p_{e\perp}}}
\newcommand{\pze}{{p_{e\parallel}^{\mbox{\protect\scriptsize edge}} }}
\newcommand{\ppe}{{p_{e\perp}^{\mbox{\protect\scriptsize edge}} }}

\def\eq#1{(\ref{eq:#1})}
\def\ddt#1#2{\partial #1 /  \partial #2}
\def\dd#1#2{\fd{\partial #1}{  \partial #2}}
\newcommand{\dpzp}{{\Delta p_{e\|\perp}\sscr{max}}}

\begin{document}

\preprint{AIP/123-QED}


\title{The Force Balance of Electrons  During Kinetic 
Anti-parallel Magnetic Reconnection}



\author{J. Egedal}
\email{egedal@wisc.edu}
\author{H. Gurram}
\author{S. Greess}
\affiliation{Department of Physics, University of Wisconsin-Madison, Madison, Wisconsin 53706, USA}

\author{W. Daughton}
\author{A. L\^{e}}
\affiliation{Los Alamos National Laboratory, Los Alamos, New Mexico
87545, USA}


\date{\today}

\begin{abstract}
Fully kinetic simulations are applied  to the study of 2D anti-parallel reconnection, elucidating the dynamics by which the electron fluid maintains force balance within both the  ion diffusion region (IDR) and the electron diffusion region (EDR). Inside the IDR, magnetic field-aligned electron pressure anisotropy ($\pz\gg\pp)$ develops upstream of the EDR. 
Compared to previous investigations, the use of modern computer facilities allows for simulations at the natural proton to electron mass ratio $m_i/m_e=1836$. In this high-$m_i/m_e$-limit the electron dynamics changes qualitatively, as the electron inflow to the EDR is enhanced and  mainly driven by the anisotropic pressure.  Using a coordinate system with the $x$-direction aligned with the reconnecting magnetic field and the $y$-direction aligned with the central current layer, it is well-known that for the much studied 2D laminar anti-parallel and symmetric scenario the reconnection electric field at the $X$-line must be balanced by the 
$\ddt{p_{exy}}{x}$ and $\ddt{p_{eyz}}{z}$ off-diagonal electron pressure stress components. We find that the electron anisotropy upstream of the EDR  imposes  large values of  $\ddt{p_{exy}}{x}$ within the EDR, and along the direction of the reconnection $X$-line this  stress cancels with the stress of a previously determined theoretical form for $\ddt{p_{eyz}}{z}$. The electron frozen-in law is instead broken by  pressure tensor gradients related to the direct heating of the electrons by the reconnection electric field. The reconnection rate is free to adjust to the value imposed externally by the plasma dynamics at larger scales. 
\end{abstract}


\maketitle
    
\section{Introduction}

Magnetic reconnection \cite{dungey:1953} 
is the process that permits the magnetic topology of an electrically conducting plasma to rearrange, often accompanied by large scale release of stored magnetic energy. 
Well-known  instances of reconnection occur in association with solar flares \cite{masuda:1994} and magnetic storms in the Earth's magnetosphere \cite{mcpherron:1973}. 
In resistive MHD models, reconnection is a slow diffusive process within narrow current layers \cite{parker:1957}. 
In low collisional plasma, however, additional physical effects become important as the theoretical width of the resistive current layers fall below the ion skin-depth $d_i=c/\omega_{pi}$. 
 Here $c$ is the speed of light and $\omega_{pi}$ is the ion plasma frequency. 
In these narrow current layers the ions' inertia may cause them to decouple from the motion of the magnetic field which has been shown to have dramatic implications for the reconnection process. 

To more accurately describe the narrow current layers of reconnection, two-fluid models treat the ions and the electrons as separate fluids and successfully predict both fast reconnection and the basic structure of the reconnection region observed in more advanced kinetic models \cite{birn:2001}. The two-fluid reconnection geometry includes an ion diffusion region (IDR) with the width of about $1d_i$, and within this region the ion frozen-in condition is broken. Mathematically, this occurs when  ${\bf E} + {\bf u}_i\times\B\ne 0$, where ${\bf E}$ and $\B$ denote the electric and magnetic fields, and ${\bf u}_i$ is the ion bulk flow velocity. Meanwhile, the electrons with bulk flow ${\bf u}_e$) remain frozen to the field line motion obeying the electron frozen-in law (${\bf E} + {\bf u}_e\times\B= 0$) almost all the way into the topological X-line. Ultimately, the frozen-in law is also broken for the electrons within the much smaller electron diffusion region (EDR) with a width of the order of the electron skin-depth, $d_e=c/\omega_{pe}$, where $\omega_{pe}$ is the electron plasma frequency. Numerical fluid models which fully resolve $d_e$ often break the electron frozen-in law by adding an ad hoc amount of hyper-resistivity (scaling in strength with the Laplacian of the plasma current density, $\nabla^2 J$) \cite{ohia:2012}. An important result of this type of investigations was the realization that the rate of reconnection becomes insensitive to the physics that are responsible for decoupling the electrons from the motion of the magnetic field lines within the EDR \cite{shay:1999}.

Despite substantial efforts including laboratory, spacecraft, numerical and theoretical investigations \cite{zweibel:2016}, a complete understanding of the electron dynamics within the EDR is still not fully developed, and its study is complicated by the process being sensitive to a range of parameters describing a given configuration. 
Open questions are also concerned with the extent to which 3D effects fundamentally alter the structure of the reconnection region \cite{silin:2005,che:2011,munoz:2017}. 
However, observations from the Magnetospheric Multiscale (MMS) mission demonstrate that the widths of reconnection layers typically approach the small length scales associated with the local electron orbit width. The length scales perpendicular to a particular reconnection current layer are  usually much shorter than the length scale along the current layer, and a range of observations further suggest that the local reconnection dynamics  are well captured by 2D laminar models \cite{egedal:2018,egedal:2019,greess:2021,schroeder:2022}.

The present paper is focused on the case of 2D symmetric and anti-parallel reconnection, which observations suggest is applicable to reconnection in the  Earth's magnetotail. Here, the opposing inflow regions often have similar strength magnetic fields but are oppositely directed \cite{torbert:2018}. Our numerical investigation also applies the typical ion to electron temperature ratio, $T_i/T_e=5$, of the magnetosphere (see Ref.~\cite{xuanye:2020}, Fig.4);  the results are relatively insensitive to this ratio. Meanwhile, just a small guide-magnetic field in the direction of the current layer can significantly alter the reconnection process, and the present results (obtained with $B_g=0$) only apply to cases where $B_g \leq 0.05 B\scr{rec}$, corresponding to Regime I reconnection as introduced in Ref.~\cite{le:2013}. While not explicitly imposed, the number densities of the electrons and ions are always similar,  consistent with the condition of quasi-neutrality being  well satisfied in the simulations. Thus, throughout the paper we will use $n_e= n_i \equiv n$.

Within the confines of the closed field lines of the plasma sheet in the Earth's magnetotail,  the normalized electron pressure, $\beta_{e}= nT_{e}/(B/(2\mu_0))\simeq 0.1$ is typical \cite{xuanye:2020}. However, if reconnection persists for a sufficient duration, low-$\beta_{e}$ plasma from the lobe regions outside the plasma sheet can reach the reconnection inflows. This can cause a dramatic drop in the plasma inflow density, reducing the electron pressure  such that $\beta_{e} \ll 0.1$. For such low values of $\beta_{e}$, the reconnection dynamics change dramatically with the generation of strong electrostatic turbulence including colliding electron holes and double layers \cite{egedal:2012,egedal:2015}. In the present paper, the numerical simulations cover the range $0.008\leq \beta_e \leq 0.5$,  and the present work is thus limited to the conditions typically observed during symmetric reconnection in the plasma sheet.

For laminar and near steady state reconnection the force balance of the electron fluid is described by the  generalized Ohm's law, which similar to the Navier-Stokes equation for a regular fluid includes pressure and inertial effects:
\begin{equation}
\hblabel{eq:Ohmslaw}
{\bf E} + {\bf u}_e\times\B + \fd{1}{ne}\nabla\cdot {\bf p}_e+\fd{m_e}{e}{\bf u}_e\cdot \nabla {\bf u}_e=0\quad.
\end{equation}
Here the electron pressure tensor is computed by an integral over the electron phase-space density distribution,  ${\bf p}_e=m_e \int ({\bf v}_e - {\bf u}_e)^2 f_e({\bf v}_e) d^3 v_e$, with ${\bf v}_e$ representing the electron velocity.
Using a range of kinetic simulations, in the present paper we will discuss how the various terms of Eq.~\eq{Ohmslaw} become important  within separate spatial areas of the IDR and EDR. 
Compared to similar investigations a decade or more ago \cite{hesse:1999,daughton:2006,drake:2008}, a main difference is that modern computing facilities now permit routine fully kinetic simulations at the full proton to electron mass ratio, $m_i/m_e=1836$. The present analysis reveals that for simulations where this ratio  approaches its natural value, ${\bf E} + {\bf u}_e\times\B\simeq 0$ does not apply to the full IDR. The reconnection electric field, $E\scr{rec}$, is the electric field along the topological $X$-line at the center of the EDR, which by Faraday's law defines the rate at which magnetic flux it reconnected. For the inner part of the IDR (but still outside what is traditionally considered the EDR \cite{shay:2007,karimabadi:2007})  we find that the  $|E\scr{rec}|\ll |\nabla\cdot {\bf p}_e/(ne)|$ such that ${\bf u}_e\times\B \simeq - \nabla\cdot {\bf p}_e/(ne)$ becomes the main driver of the Hall magnetic field perturbation. In fact, the familiar assumption of $u\scr{in}= E\scr{rec}/B\scr{rec,up}$ significantly underestimates the inflow speed of the electron fluid into the EDR. Here, $B\scr{rec,up}$ is the in-plane magnetic field just upstream of the EDR. 

The main topic of the paper, however, is the development of a theory to explain the terms of Eq.~\eq{Ohmslaw} which break the electron frozen-in condition at the very center of the EDR.
The study emphasizes the importance of the electron anisotropy upstream of the EDR in driving the current within the EDR and cancelling an off-diagonal stress term identified in previous work \cite{kuznetsova1998kinetic,hesse:1999}. Ultimately, the off-diagonal stress of ${\bf p}_e$, that is  responsible for breaking the frozen-in condition, is generated by $E\scr{rec}$ itself.
Thus, the theory suggests that the electron fluid does not represent an obstacle (or bottleneck) for reconnection, which may then proceed at the rate imposed by dynamics external to the EDR.  In turn, once reconnection has started, its rate will be controlled by larger scale dynamics which in many cases is well described by MHD models \cite{cassak:2017,liu:2017}. In certain cases including island coalescence it can also be important to retain effects of ion pressure anisotropy  \cite{le:2014,stanier:2015}.

The paper is organized as follows: In Section II we discuss previous results on the formation and role of electron pressure anisotropy within the IDR inflow regions, and how it drives the current of the EDR. Section III explores the structure and energy balance of electron flows within the inner part of the IDR, whereas Section IV examines the length scales characterizing the EDR. A detailed account of the electron momentum balance at the $X$-line region is provided in Section V, and the paper is summarized and concluded in Section VI.

\section{Summary of previous results on the role of electron pressure anisotropy }
Our analysis relies strongly on  previous results regarding the overall force balance of the EDR \cite{le:2010grl,egedal:2013,montag:2020}. 
For the convenience of the reader,  in this section we provide a short summary of this previous work on how the upstream electron pressure anisotropy impacts the structure of the EDR. Additionally, in Appendices A and B we provide additional mathematical considerations to   demonstrate how the upstream electron pressure anisotropy drives the current jets within the EDR. 

\subsection{Formation of electron pressure anisotropy in reconnection inflows, upstream of the EDR} 

\begin{figure}[h]
	\centering
	\includegraphics[width=8.6cm]{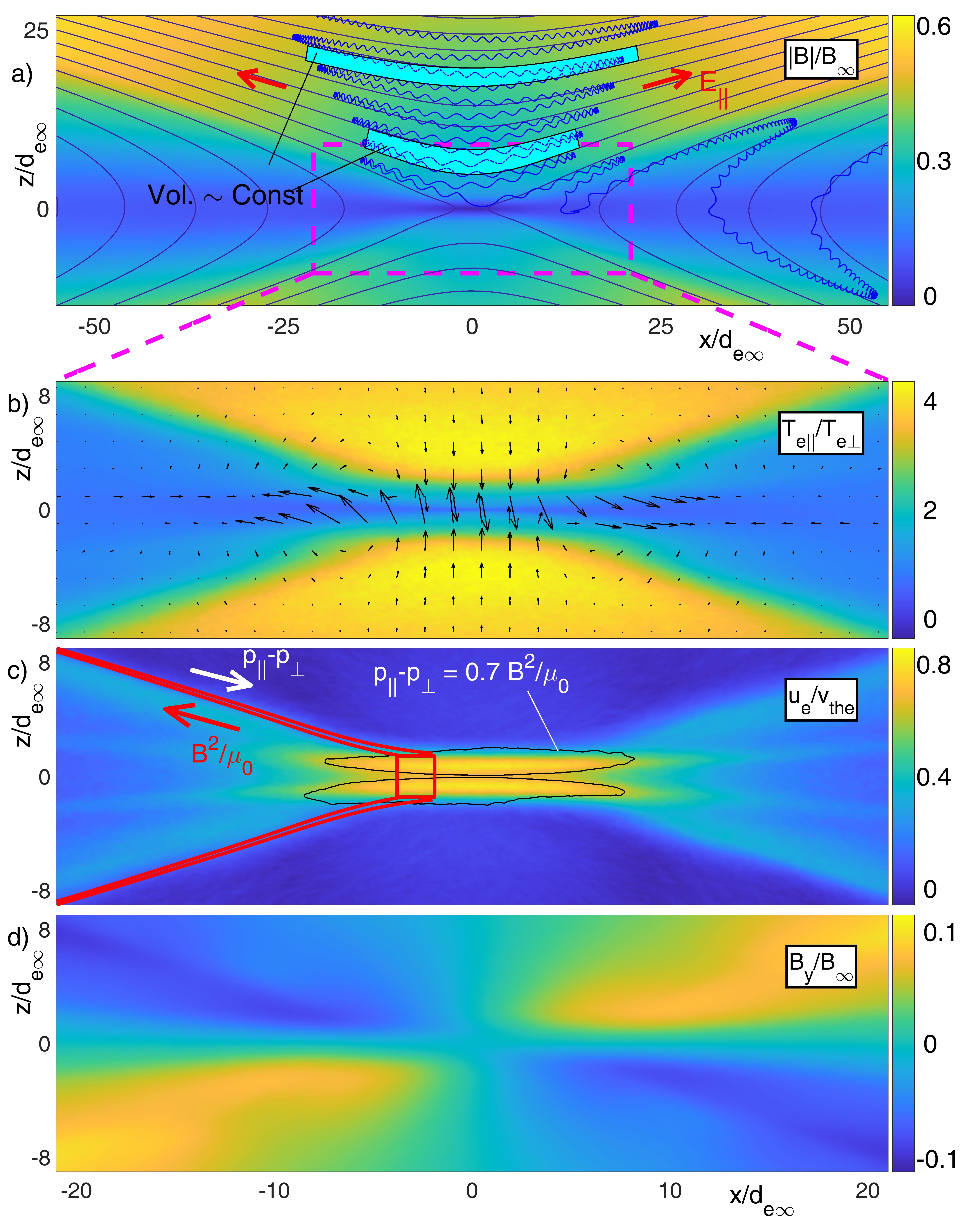}%
	\caption{VPIC simulations profiles   of a) $B/B_{\infty}$, b) $T_{e\|}/T_{e\perp}$, c) $|{\bf u}_e|/v_{te}$, and d) $B_y/B_{\infty}$, obtained with  $m_i/m_e=1836$ and $\beta_{e\infty}=2^{-4}=0.0625$. The arrows in b) represent the enhanced electron flow provided by  ${\bf J}\scr{extra} = [(\pz-\pp)/B] {\bf b}\times \mathbf{\kappa}$. }
	\vspace{-0.4cm}
	\hblabel{fig:Fig1}
\end{figure}

Due to the trapped orbit dynamics illustrated in Fig.~\ref{fig:Fig1}(a), it is generally found that the reconnection inflows are characterized by a regime of double adiabatic electron dynamics. It can be shown \cite{le:2009,le:2010grl} that at the upstream edge of the EDR the main electron pressure components are well described by the 
CGL-limit \cite{chew:1956} with $\pz\propto n^3/B^2$ and $\pp\propto nB$,  where $\|$ and $\perp$ denote the directions parallel and perpendicular relative to the local direction of  $\B$. For  $ \beta_{e\infty}^2m_i/m_e  \lesssim1$
electron holes and double layers yield even stronger $\pz$ heating \cite{egedal:2013,egedal:2015}.  The effects of the anisotropic heating  are evident in the profile in Fig.~\ref{fig:Fig1}(b), where $T_{e\|}/T_{e\perp}\simeq4$ is observed just upstream of the EDR, and the marginal electron firehose condition is approached, $\pz-\pp \simeq B^2/\mu_0$.  Within the EDR the adiabatic invariance of the electron magnetic moments, $\mu=m\vp^2/(2B)$, break, leading to isotropization and pitch angle mixing. As a result, the electron pressure is approximately isotropic   in the  reconnection exhaust \cite{le:2013,egedal:2016b}. 

As described in Appendix A, the physical mechanisms which underpin the CGL-scalings  can be understood through relatively simple arguments applicable to the electrons just upstream of the EDR, where the described trapping  and parallel compression dynamics  (see compressed trapped orbit in Fig.~\ref{fig:Fig1}(a)) are strong enough to  dominate the properties of the electron fluid.  
A more accurate and rigorous model is provided in \cite{le:2009,egedal:2013} taking into account that not all electrons become trapped. For the present paper, however, it is sufficient to assume that within the inflow regions the electron pressures approximately follow the aforementioned CGL-scaling laws.
While the present simulations use open boundary conditions, the same trapped electron dynamics occur in simulations with  periodic boundary conditions \cite{egedal:2009pop}. In contrast to periodic conditions, however, the open system do not need to accommodate the large island of reconnected flux which gets stuck in the periodic reconnection exhausts. Thus, with open boundary conditions a larger fraction of the system is available for the reconnection dynamics \cite{gurram:2021}. 
\\[1ex]

\subsection{Force balance along the length of the EDR and associated scaling laws for the electron pressure anisotropy}

In Refs.~\cite{le:2010grl,le:2010pop} it was demonstrated that the two dominant terms of Eq.~\eq{Ohmslaw} along the EDR  are ${\bf u}_e\times\B$ and $(\nabla\cdot {\bf p}_e)/ne$. Expressed in terms of the  Maxwell stress tensor, ${\bf T}= (\B\B - B^2 {\bf I})/(2\mu_0)$, we have $-e n {\bf u}_e\times\B \simeq {\bf J} \times \B=\nabla\cdot {\bf T}$, and momentum balance then requires that $\nabla\cdot ({\bf p}_e-{\bf T})\simeq0$. Considering the area encircled by the red square in Fig.~\ref{fig:Fig1}(c), it can then be shown by the divergence theorem \cite{le:2010grl,montag:2020} that at the upstream edge of the EDR the electron firehose condition must be approximately satisfied,
\begin{equation}
\hblabel{eq:fire}
\pze-\ppe \simeq B_H^2/\mu_0\quad,
\end{equation}
where $B_H$ represents the  magnetic field strength just upstream of the EDR (the direction of this field rotates along the upstream edge of the EDR into the Hall magnetic field explaining the subscript $H$). For example, for the red square in Fig.~\ref{fig:Fig1}(c)  the magnetic tension acting on the sides of the fluid element (corresponding to the integral over the element of the $-en{\bf u}_e\times\B$ force) is offset by pressure anisotropy. Again, this is discussed in more detail in Appendices A and B.

We can derive important scaling laws for the level of pressure anisotropy for locations just upstream of the EDR.
Using that $\pze\gg\ppe$ with the CGL-limits  the maximal pressure differential
\begin{equation}
\hblabel{eq:delp}
\dpzp\equiv \mbox{max}(\pz-\pp)=\pze-\ppe\quad,
\end{equation}
adheres to  the asymptotic scaling $\dpzp \simeq \pze\simeq (\pi\tilde{n}^3/(6\tilde{B}_H^2)) \beta_{e\infty} B_{\infty}^2/(2\mu_0)$.  The factor $\pi/6$ is part of the L\^e 2009 Equations of State \cite{le:2009},  and the $\infty$ subscripts refer to quantities being evaluated far upstream of the IDR, where the electron pressure is assumed to be isotropic. 
The firehose condition in Eq.~\eq{fire}, representing the dominant momentum balance condition of the EDR, may then be expressed as 
\begin{equation}
\hblabel{eq:BH}
\fd{B_H}{B_{\infty}} = \fd{\mu_0 K}{2B_{\infty}} = \left(\fd{\pi \tilde{n}^3 \beta_{e\infty}}{12}\right)^{1/4}
\quad.
\end{equation}
 Here $K$ is the integral of the current density across the layer, representing the current per unit length of the EDR. $\tilde{n}\simeq 1$ is the local electron density normalized by the upstream value $n_{\infty}$ (and similar for $\tilde{B}_H$ used above) \cite{le:2010grl}. 
Eq.~\eq{BH} is significant as it demonstrates how $K$ is mainly controlled by the external electron pressure anisotropy, which through the CGL-like scalings is set by the normalized upstream electron pressure, $\beta_{e\infty}$.

The above result is in contrast to predictions  of two-fluid models with isotropic pressure where ${\bf u}_e\times\B$ is balanced by resistive terms, ${\bf E}$,  and/or $(m_e/e){\bf u}_e\cdot \nabla {\bf u}_e$.
 As discussed below, the half-width  $l_z\simeq 2 d_{e\infty}$ of the EDR region current layer is set by the meandering orbit width \cite{horiuchi:1997,shay:1998, roytershteyn:2013}. From Amp\`{e}re's law it then follows that $B_H= \mu_0 e n u_e l_z \simeq 2\mu_0 e n u_e d_{e\infty}$, such that $u_e = B_H/(2en u_ed_{e\infty}) \simeq   B_H/(2\sqrt{\mu_0n m_e}) = v_{Ae}/2$. Thus,
 $u_e$  must approach the corresponding value of $\sim v_{Ae}/2$, simply because of the narrow width of the current layer $l_z\simeq 2 d_{e\infty}$.  This analysis is also applicable to the Harris sheet type current layer and the condition $u_e = v_{Ae}/2$ is therefore reflective of the underlying meandering orbit dynamics setting the narrow width of the electron flow region \cite{roytershteyn:2013}. 
The EDR  current layer direction rotates as a function of $x$ in the $xy$-plane \cite{mandt:1994,hesse:2008,le:2010pop}, and  yields the $B_y$ perturbations shown in  Fig.~\ref{fig:Fig1}(d) \cite{hesse:2008,le:2010grl}. The above results have also been applied in deriving scaling laws for the absolute electron heating within the reconnection inflow and EDR \cite{le:2016twostage}.

\section{Electron flows and energy balance within the inner part of the IDR}

\subsection{Setup of kinetic simulations}

The  2D kinetic simulations applied in our study were implemented for  anti-parallel Harris
sheet reconnection using the code VPIC \cite{bowers:2009} with open boundary
conditions \cite{daughton:2006}.  The initial magnetic field and
the density are $B_x=B_0 \tanh(z/\lambda)$ and $n=n_0\cosh^2(z/\lambda)+n_{\infty}$, respectively, where $\lambda= 0.5d_{i0}=0.5\sqrt{\varepsilon_0 m_i c^2/n_0e^2}$.  Other parameters are initial uniform temperatures with $T_{i\infty}/T_{e\infty}=5$, $\omega_{pe0}/\omega_{ce0}=2$, and $\sim400$ particles per species per cell. The study incorporates 35 separate simulation runs implementing a matrix of 5 values for $m_i/m_e$ and 7 values for the normalized upstream electron pressure, $\beta_{e\infty}$. 
To be more specific, we use mass ratios of $m_i/m_e\in\{ 100, 200, 400, 800, 1836\}$, and the background density $n_{\infty}$ is varied so the initial upstream electron beta is
$\beta_{e\infty}=2\mu_0 n_{\infty} T_{e\infty}/B_0^2=2^{-k}$ with $k\in \{1, 2, \dots, 7\}$. In VPIC ``natural'' units the systems are implemented using $m_e=1$, $d_{e0}=1$, $c=1$, $n_0=1$, $B_0=1/2$, $T_{e\infty}= (1/48)\, m_ec^2$, and $n_{\infty}/n_0 = (1+T_{i\infty}/T_{e\infty})2^{-k}$.  

For runs with $m_i/m_e \leq 200$ the domains were $3960\times1980$ cells, corresponding to 
$ 50 d_{ip}\times 25 d_{ip}$. Runs at 
$m_i/m_e\geq 400 $ were carried with $5632\times2816$ cells, corresponding to 
system sizes of  $1000 d_{ep}\times 500 d_{ep}$.
Here, 
$d_{ip}$ and $d_{ep}$ are the  ion and electron skin-depths, respectively, defined with respect to the central peak Harris sheet density, $n_p= n_0+n_{\infty}$. Furthermore, the upstream electron skin-depth becomes central to our analysis below and can be expressed as $d_{e\infty}=d_{e0}/\sqrt{n_{\infty}/n_0}$. 
For all of the runs,  a fixed electron temperature is applied such that $v_{te}/c = 0.144  = L\scr{Debye} /d_{e0}$. Measuring the cell size, $\delta$, relative to the Debye length, we then have  $\delta/ L\scr{Debye} \simeq  (1000/5632) \cdot 0.144 \simeq 1.2$, again, based on the  initial temperature. The computational cost of the runs scales approximately as $(m_i/m_e)(L_x/d_{ep})^2$.
More details about the similarities and differences between the individual simulation runs are provided in Section IV, where the lengths  and widths of the EDR electron jets are examined. In addition, in Table~\ref{tab:table1} of Appendix D we provide the simulation domain sizes normalized by various parameters.

\subsection{The reconnection rate from the perspective of the electrons}

Consistent with external MHD scale constraints  for system sizes larger than $10d_{i\infty}$ \cite{liu:2017}, in the present simulations the absolute reconnection rate obeys  $E\scr{rec}\simeq \alpha v_A B$. Here  $\alpha\simeq 0.1$ is the normalized reconnection rate, and the relevant Alfv\'en speed and $B$ are evaluated $1d_i$ upstream of the EDR \cite{shay:2004}. 
 In our study, values of $E\scr{rec}$ are computed through a spatial average of $E_y$ over a $2d_{e\infty}\times 0.5 d_{e\infty}$ region centered on the topological $X$-line. 
In Fig.~\ref{fig:Plots3mod10}(a),
the corresponding  values of $\alpha$ are shown for the full range of our simulations. It should be noted that for $m_i/m_e=1836$ and $k=7$ the numerical domain only measures about $5d_{i\infty}\times 2.5d_{i\infty}$ (see Table I.C in Appendix D). This could explain the slightly enhanced normalized reconnection rate of $\alpha\simeq 0.14$, which is expected as the ``electron-only'' regime is approached at small system sizes \cite{phan:2018,pyakurel:2019,olson:2021,greess:2022}. Nevertheless, other runs in the set show similar values of $\alpha$ and, as will become clear below,  the dynamics recorded for this most extreme run still conform well with the general trends of the full data set.

In Section IV we document how the size of the EDR normalized by $d_{e\infty}$ remains approximately constant for varying values of $m_i/m_e$. Meanwhile, the size of the IDR scales with $d_{i\infty} =  d_{e\infty} \sqrt{m_i/m_e} $, and from the perspective of the electrons the size of the ion diffusion region thus increases by a factor of $\sqrt{m_i/m_e}$. Likewise, relative to the time scale of the electron motion, the increasing inertia of the ions will slow the rate of reconnection. As a dimensionless measure of the reconnection electric field relevant to the electron orbit dynamics we introduce 
  $\hat{E}\scr{rec} = e d_{e\infty} E\scr{rec}/T_{e\infty}$. This quantity represents the temperature-normalized energy gain an electron will acquire when traveling $d_{e\infty}$ in the direction of $E\scr{rec}$. 
Corresponding to the data shown in Fig.~\ref{fig:Plots3mod10}(b), it then follows that $\hat{E}\scr{rec} \propto \sqrt{m_e/(m_i\beta_{e\infty}^2)}$, where reduced values of $m_i/m_e$ and low $\beta_{e\infty}$ impose the largest values of $\hat{E}\scr{rec}$. 

In Fig.~\ref{fig:Plots3mod10}(c) we show the profiles of $T_{e\|}/T_{e\infty}$ and $T_{e\perp}/T_{e\infty}$ corresponding to values in the various runs recorded just upstream of the electron diffusion region. Consistent with the discussion in Section II, for simulations within the double adiabatic regime ($\beta_{e\infty}^2m_i/m_e\gtrsim 1$) marked by the full lines, we observe that both of these profiles are largely independent of $m_i/m_e$. The normalized electron pressure anisotropy $(p_{e\|}-p_{e\perp})/(n_\infty T_{e\infty})$ is then also independent of $m_i/m_e$, and given the scaling of $\hat{E}\scr{rec} \propto \sqrt{m_e/m_i}$ the forces associated with the pressure anisotropy  become most significant when compared to $e\hat{E}\scr{rec}$ at large values of $m_i/m_e$. In fact, as 
illustrated in Fig.~\ref{fig:Plots3mod10}(d) and to be discussed further below, the thermal forces of the electron pressure anisotropy dominate the force balance of the electrons for $m_i/m_e=1836$, rendering the electron dynamics of the IDR and EDR significantly different when compared to reduced fluid models  invoking isotropic electron pressure. 

\begin{figure}[h]
	\centering
	\includegraphics[width=8.4 cm]{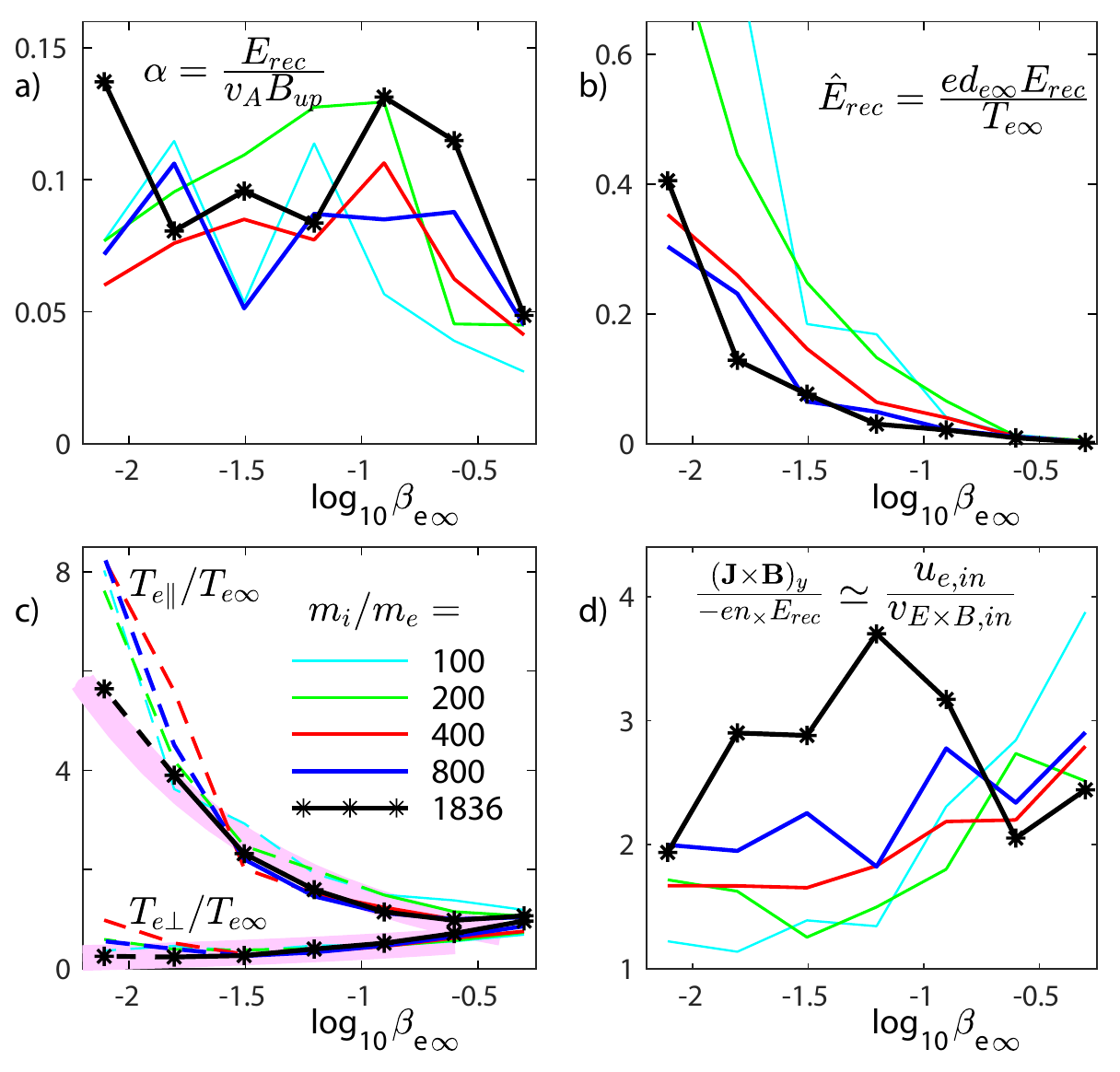}%
	\caption{  Obtained from a matrix of 35 kinetic simulations, profiles as functions of $\log_{10}(\beta_{e\infty})$ are shown for a range of parameters as marked in the panels. As indicated in c), the line-type denotes the value of $m_i/m_e$. 
	The ``ion-normalized'' reconnection rate $\alpha = E\scr{rec}/(v_A B_{up})$ is shown on a)
 while in b) $\hat{E}\scr{rec}= e d_{e\infty} E\scr{rec}/T_{e\infty}$ is the normalized  rate relevant to the electron dynamics. 
		In c) the regions shaded in magenta are obtained from the CGL-limit of the L\^e-2009 Equations of State \cite{le:2009}. The dashed lines  mark runs in the regime $\beta_{e\infty} \lesssim \sqrt{m_i/m_e}$ where additional $T_{e\|}$ heating is typically observed \cite{egedal:2015,le:2016twostage}. The profiles in d) show the maximum value of $({\bf J}\times{\B})_y$ observed just upstream of the EDR, normalized by $-en_{\times} E\scr{rec}$, documenting how forces related to ${\bf J}\times{\B}$ and $\nabla\cdot{\bf p}_e$ dominate at the natural value of $m_i/m_e$.
	}
	\hblabel{fig:Plots3mod10}
\end{figure}

\subsection{Dominant role of $p_{e\|} \gg p_{e\perp}$  in setting the electron flows and energy dissipation upstream of the EDR}
\begin{figure}[h]
	\centering
	\includegraphics[width=0.47\textwidth]{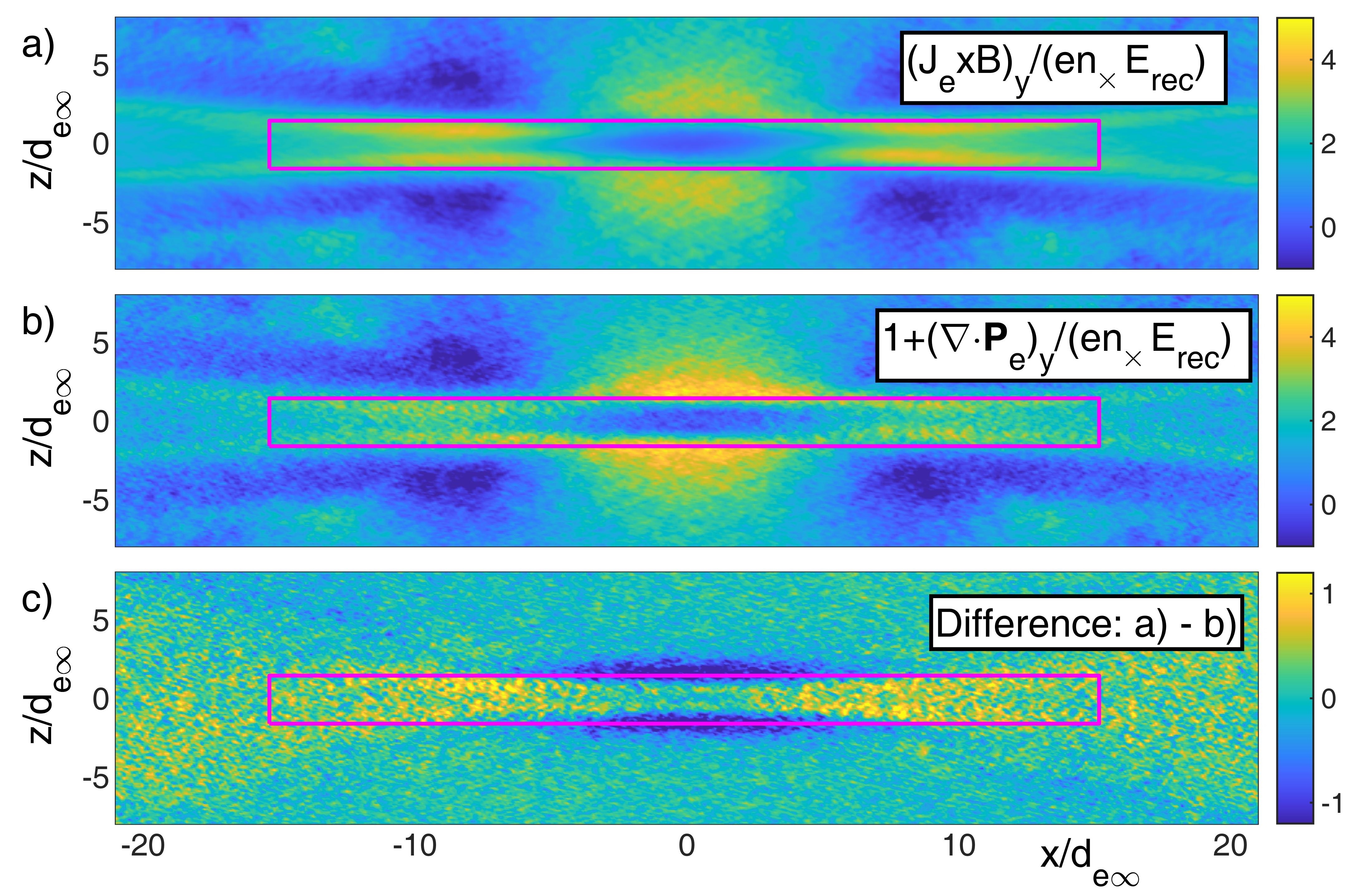}%
	\caption{Considering the same simulation as in Fig.~\ref{fig:Fig1} ($m_i/m_e=1836$ and $\beta_{e\infty}=2^{-4}$), color contours are presented for a) the LHS and b) RHS of Eq.~\eq{OhmsLawMod2}. The EDR is outlined by the magenta rectangles. 
 For the extended regions where $({\bf J}\times\B)_y/(n_{\times} E\scr{rec}) \gtrsim 3.5$ the electron fluid is moving faster than the magnetic field. This motion is driven by electron pressure anisotropy such that $|{\bf J}\times\B| \simeq |\nabla \cdot {\bf p}_e| \gg e n_{\times} E\scr{rec}$. The difference between panels a) and b) is shown in c) and can be attributed to  the neglected electron inertia term (note the change in color scale).
	}
	\hblabel{fig:Fig1sup2mod}
\end{figure}

In the region outside the EDR where the electrons are well magnetized,  
the electron pressure anisotropy ($\pz\gg\pp$) drives additional currents, ${\bf J}\scr{extra} = [(\pz-\pp)/B] {\bf b}\times \mathbf{\kappa}$, via the curvature drift; these currents are not observed for isotropic pressure \cite{egedal:2013}. Here 
$\mathbf{\kappa}=({\bf b}\cdot \nabla) {\bf b}$ is the curvature vector, which becomes large where the magnetic field direction, ${\bf b}=\B/B$, has strong spatial variation. In Fig.~\ref{fig:Fig1}(b) the arrows are proportional to 
$-{\bf J}\scr{extra}$ corresponding to a flow boosting the perpendicular drift of electrons into the EDR.   ${\bf J}\scr{extra}$ also has a component in the $y$-direction which drives currents outside the EDR that are in the opposite direction of the electron current within the EDR. While the mathematical expression for ${\bf J}\scr{extra}$ is not valid inside the EDR, as discussed in Section II the out-of-plane electron current of the EDR can still be attributed to similar effects directly related to  $\pz\gg\pp$.  

To quantify the relative importance of the flow by  ${\bf J}\scr{extra}$, in Fig.~\ref{fig:Fig1sup2mod} we provide additional profiles for  the simulation of Fig.~\ref{fig:Fig1},  here focusing on the force balance in the $y$-direction in the vicinity of the EDR.  
Within the IDR we generally have $({\bf u}_e\times \B)_y \gg ({\bf u}_i\times \B)_y$ and we thus introduce the approximation
that $(-en{\bf u}_e\times \B)_y \simeq ({\bf J}\times\B)_y$.  Consistent with previous work \cite{vasyliunas:1975}, generally we have that  $\left|({\bf J}\times\B)_y \right| \gg \left| (en m_e {\bf u}_e\cdot\nabla  {\bf u}_e)_y\right|$, and the $y$-component of the electron inertia can for the present analysis be neglected. With these approximations the $y$-component of the generalized  Ohm's law in Eq.~\eq{Ohmslaw}  then becomes \cite{cai:1994} 
\begin{equation}
    \hblabel{eq:OhmsLawMod1}
    -enE_y + ({\bf J}\times\B)_y = (\nabla\cdot {\bf p}_e)_y\quad.
  \end{equation}
 For the time point considered in Fig.~\ref{fig:Fig1sup2mod} the reconnection geometry and rate is approximately steady. Furthermore,  the density profile in this inner region is approximately uniform at a density we denote as  $n_{\times}$. It then follows that $n E_y \simeq n_{\times} E\scr{rec}$, where $E\scr{rec}$ is the value of $E_y$ observed at the $X$-line. After dividing Eq.~\eq{OhmsLawMod1}  by $en_{\times} E\scr{rec}$ we obtain:
 \begin{equation}
    \hblabel{eq:OhmsLawMod2}
    \fd{({\bf J}\times\B)_y}{en_{\times}E\scr{rec}} \simeq 1+ \fd{(\nabla\cdot {\bf p}_e)_y}{en_{\times}E\scr{rec}}\quad.
  \end{equation}
The LHS (left hand side) of Eq.~\eq{OhmsLawMod2} is shown in Fig.~\ref{fig:Fig1sup2mod}(a), while the RHS is shown in  Fig.~\ref{fig:Fig1sup2mod}(b).  The difference between the profiles is relatively small and is shown in
Fig.~\ref{fig:Fig1sup2mod}(c). This difference can be attributed to the neglected electron inertia in Eq.~\eq{OhmsLawMod2}. 
 
From Fig.~\ref{fig:Fig1sup2mod}(a) it is apparent  that in the vicinity of the EDR $({\bf J}\times\B)_y$ is significantly larger than  $en_{\times}E\scr{rec}$.
Thus, rather than $en E_y \simeq ({\bf J}\times\B)_y$,  we see that
$|enE_y| \ll |({\bf J}\times\B)_y|\simeq |(\nabla\cdot {\bf p}_e)_y|$ in the important inner part of the IDR and throughout most of the EDR. 
Within the IDR for $x=0$, it is readily shown that $({\bf J}\times\B)_y / 
en_{\times}E\scr{rec} = u_{e,z}/v_{E\times B,z}$, and the  enhanced inflow is provided by $p_{e\|}\gg p_{e\perp}$ through the drifts associated with ${\bf J}\scr{extra}$. In other words, while $E_y$ by Faraday's law is fundamental for moving the magnetic flux across the reconnection region,  Fig.~\ref{fig:Fig1sup2mod} reveals how the $(\nabla\cdot {\bf p}_e)_y$-term is responsible for driving the electrons into the EDR and through the elongated jets of the EDR at high rates, which for the present run are up to 3.5 times the speed of the magnetic field line motion. In Fig.~\ref{fig:Plots3mod10}(d) the maximum value of $({\bf J}\times\B)_y / en_{\times}E\scr{rec}\,\, (= u_{e,z}/v_{E\times B,z}) $ is shown for each of the runs, demonstrating how the enhancement of the inflow speed becomes most pronounced as $m_i/m_e=1836$ is approached.

The  electron pressure anisotropy also has implications for the energy dissipation within the reconnection region. 
It was recently emphasized \cite{liu:2022} that an electron fluid obeying the frozen-in condition, ${\bf E} = -{\bf v}_e \times {\bf B}$, does not exchange any energy with the electromagnetic fields simply because  ${\bf J_e}\cdot {\bf E}
=en {\bf v}_e \cdot ({\bf v}_e \times {\bf B})=0$. This observation is altered by the presence of the strong electron pressure anisotropy. For this case
${\bf E} = -{\bf v}_e \times {\bf B}- \nabla \cdot {\bf p}_e/(ne)$, such that the now non-zero value of  
${\bf J_e}\cdot {\bf E} = {\bf v}_e\cdot (\nabla \cdot {\bf p}_e)$  permits energy to be exchanged with the fields. In Fig.~\ref{fig:JE}(a) we display the profile of $-{\bf J_e}\cdot {\bf E}$, where
``yellow'' areas with $-{\bf J_e}\cdot {\bf E}> 0$ are regions where the electrons give energy back to the fields. 
Such regions, known as generator regions, have been directly observed by the MMS mission \cite{payne:2021}.

The continuity of the electromagnetic energy is described by Poynting's theorem, $-\ddt{u_{EB}}{t}= \nabla \cdot {\bf S}+{\bf J}\cdot {\bf E}$, where $u_{EB}= \varepsilon_0 E^2/2 + B^2/(2\mu_0)$ is the electromagnetic field energy density and ${\bf S} = {\bf E}\times {\bf B}/\mu_0$ is the Poynting flux. The profile of  $\nabla \cdot {\bf S}$ is shown in Fig.~\ref{fig:JE}(b), which closely resembles the profile of $-{\bf J}_e\cdot {\bf E}$. The difference in Fig.~\ref{fig:JE}(c) is the energy either exchanged with  $u_{EB}$ and/or the ions through ${\bf J}_i\cdot {\bf E}$. The smallness of this difference is consistent with a steady state scenario with steady energy density of the fields throughout the region.  

 As indicated above, the identified generator regions just upstream of the EDR are a direct consequence of the electron pressure anisotropy that develops in the inflow regions. This is emphasized when comparing $-{\bf J_e}\cdot {\bf E}$ in Fig.~\ref{fig:JE}(a) with  $-{\bf J}\scr{extra}\cdot {\bf E}_z$ in Fig.~\ref{fig:JE}(d). Again, the expression for ${\bf J}\scr{extra}$ is not valid within the EDR. Nevertheless, the predicted values of ${\bf J}\scr{extra}\cdot {\bf E}_z$ 
demonstrate that the main contributor to  ${\bf J_e}\cdot {\bf E}$ at the edge of the EDR are the curvature-drift-driven currents that enhance the inflow of the electrons into the strong $E_z$ electric fields on the EDR's edge. The structure of $E_z$ will be discussed in further details below.

\begin{figure}[h]
	\centering
	\includegraphics[width=0.47\textwidth]{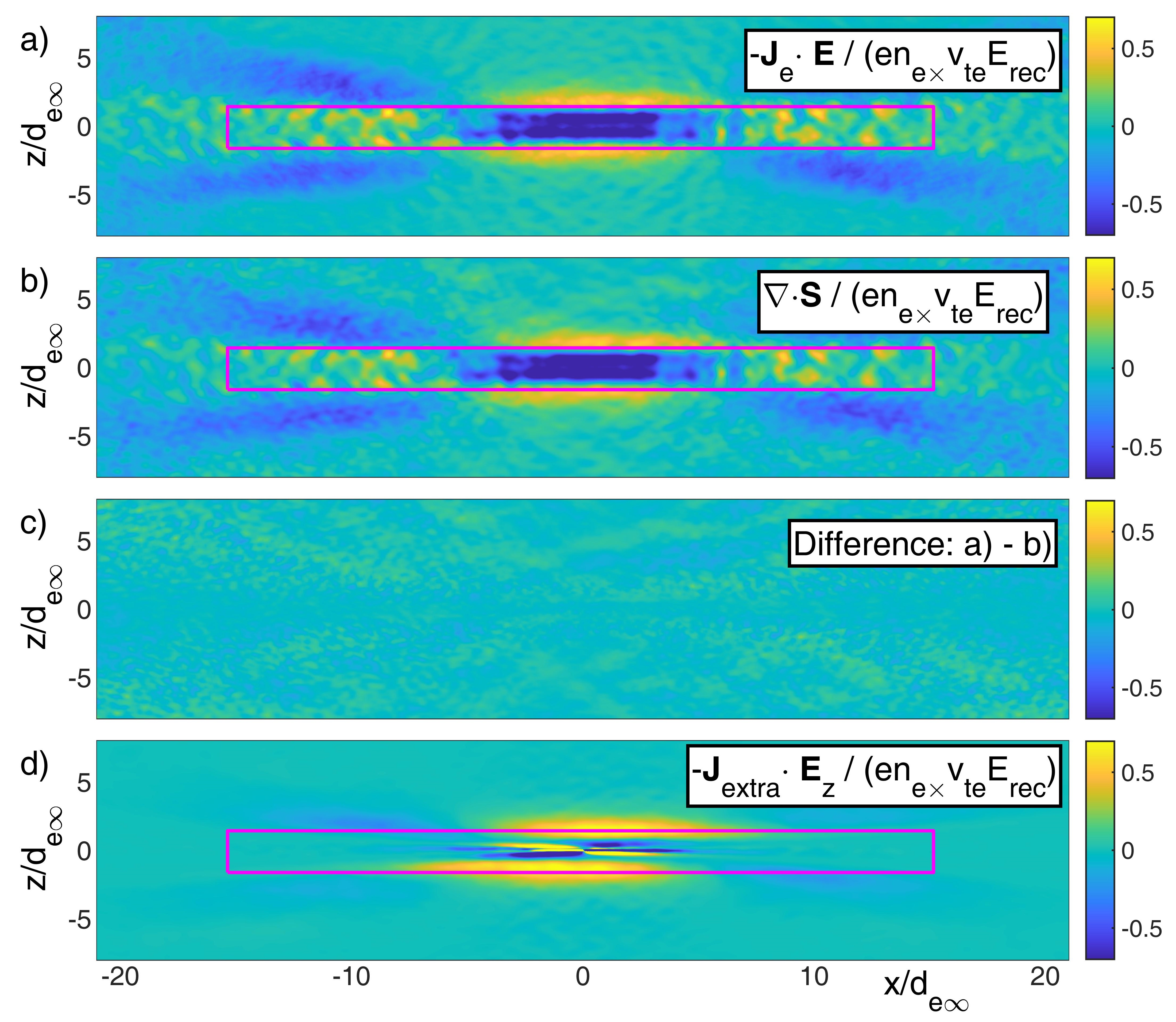}%
	\caption{a,b) Near identical profiles of $-{\bf J}_e\cdot \E$ and $\nabla\cdot {\bf S}$, representing the main two terms in Poynting's theorem. Their difference is shown in c). The profile of 
 $-{\bf J}\scr{extra}\cdot \E_z$ is shown 
 d). }
	\hblabel{fig:JE}
\end{figure}


  \begin{SCfigure*}[][h]
  	\vspace{-0.1in}
  	\includegraphics[width=13cm]{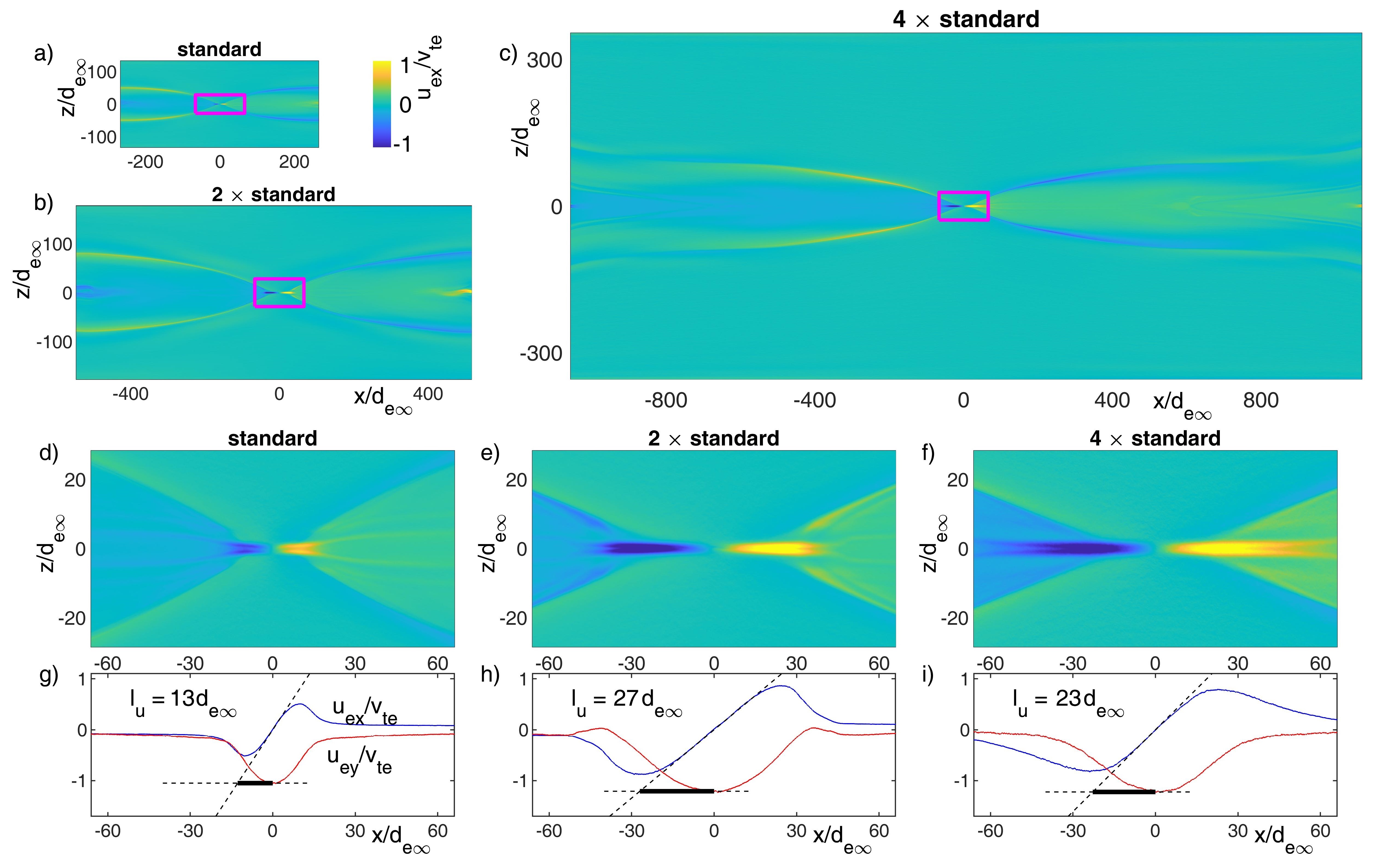}
  	 \caption{ a-c) Profiles of $u_{ex}/v_{te}$ for the full simulation domains for three separate runs at $\beta_{e\infty}=2^{-4}$ and $m_i/m_e=800$. The run in a) has the ``standard'' size described in Section II.A, whereas the runs in b,c) are for simulations with spatial domain sizes increased by factors $2\times$ and $4\times$, respectively. Zoomed-in views of the EDRs are shown in d-f), whereas g-i) show cuts of $u_{ex}$ and $u_{ey}$ for $z=0$, from which the length scale $l_u$ of current layer rotation is determined.    }
  	\hblabel{fig:do800}
  \end{SCfigure*}

\section{Length scales of the EDR}


We now introduce a new length $l_u$, which is a measure of the spatial scale at which the direction of the out-of-plane EDR electron current at the  $X$-line rotates into the exhaust direction. This length scale turns out to be fundamental our momentum balance analysis of the EDR. 


\subsection{Numerical evidence that $l_u$ does not scale with simulation system size}

In Ref.~\cite{daughton:2006}, which employed relatively low mass ratio $m_i/m_e=25$, it was observed that the length of the EDR increases with the size of the numerical simulation domain, and evidence was presented that the long electron current layers of the EDR could act as a bottleneck for reconnection. These results do not to apply to our present simulations where $m_i/m_e$
approaches its natural value of 1836. For example, in Fig.~\ref{fig:do800}(a) profiles of $u_{ex}$ are shown for a simulation at $m_i/m_e=800$ and $\beta_{e\infty}=2^{-4}$ for the ``standard'' setup described in Section II.A, whereas Figs.~\ref{fig:do800}(b,c) apply domains factors of $2\times$ and $4\times$ larger, respectively. Zoomed-in views of the EDRs for the three runs are shown in Figs.~\ref{fig:do800}(d-f).  In the standard case the length of the EDR is likely reduced due to the proximity of the simulation boundaries, whereas we notice how the $2\times$ and $4\times$ runs have longer EDR lengths. This indicates that for simulations at large $m_i/m_e$, the length of the EDR becomes independent of the system size when the run is implemented in a sufficiently large domain.

For the analytical analysis of the EDR  the following definition of $l_u$ becomes essential to the presented theory for the force balance of the electron fluid. Here
\begin{equation}
    \hblabel{eq:lu}
l_u\equiv \left| \frac{u_{ey}}{du_{ex}/dx}\right|\scr{X-line}, 
\end{equation}
represents a measure of the spatial distance that characterizes  the rotation of the EDR electron current direction in the $xy$-plane as a function of $x$.
Figs.~\ref{fig:do800}(g-i) provide the geometric interpretation of this quantity, which is observed to be on the order of $20d_e$ for the three runs. Below we show how $l_u$ is significantly shorter than the full length $l\scr{jet}$ of the EDR  for many runs in our study, consistent with 
previously observed two-scale EDR structures  \cite{karimabadi:2007,shay:2007}.

\subsection{Numerical results for $l_u$ over the full $(\beta_{e\infty}, m_i/m_e)$--matrix of kinetic runs}

 \begin{figure*}[ht]
	\centering
	\includegraphics[width=\textwidth]{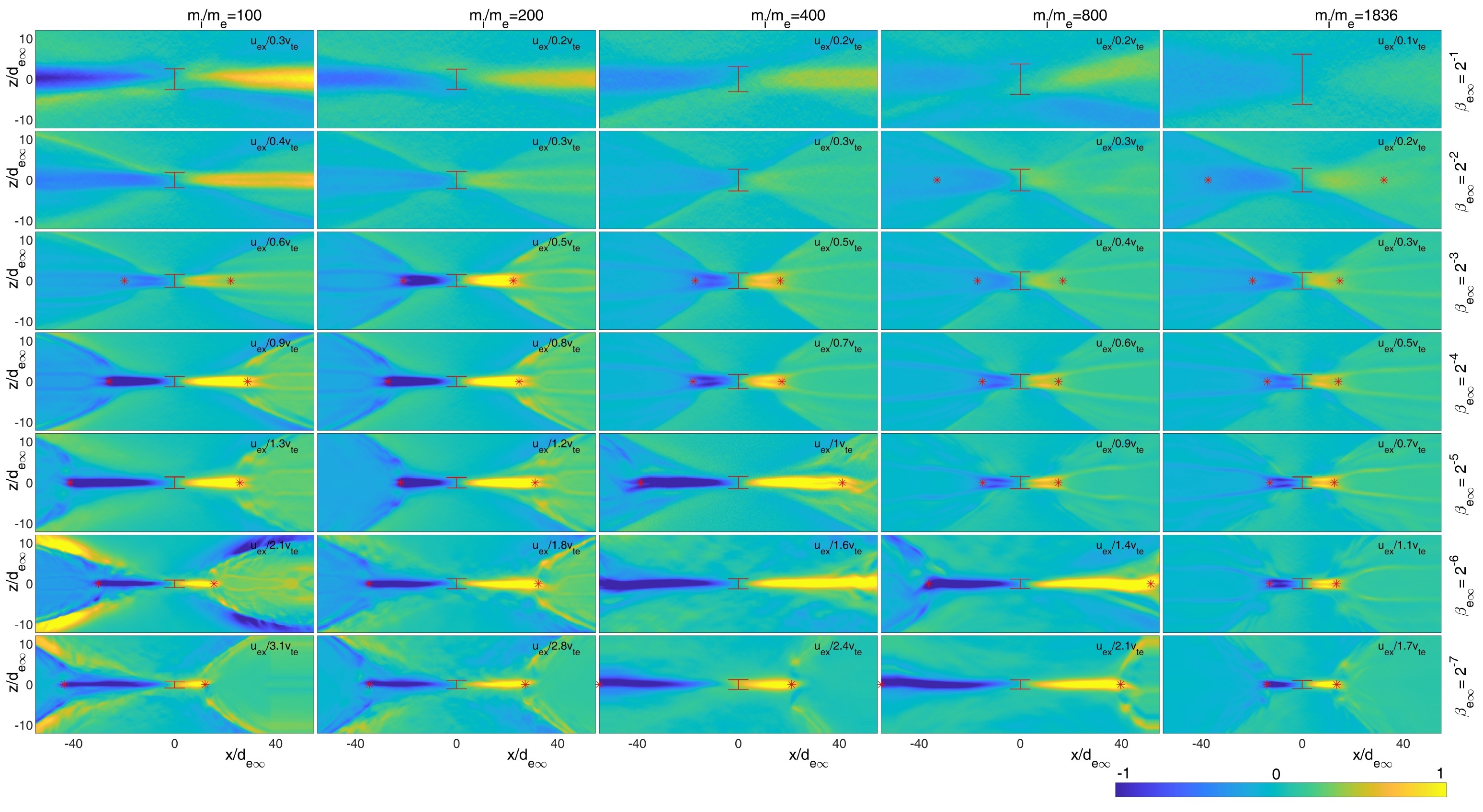}
	\caption{Color contours of $u_{ex}$ for the matrix of numerical simulations applied in the study,
		$m_i/m_e\in\{ 100, 200, 400, 800, 1836\}$ and $\beta_{e\infty}=2^{-k}$ with $k\in \{1, 2, \dots, 7\}$.
		When falling within the domains considered, the ends of the jets are marked by the red stars corresponding to $|u_{ex}|$ dropping to 60\% of its peak value. The similarly defined width of the $u_{ey}$ out-of-plane electron drift is marked in each panel at $x=0$. 
	}\hblabel{fig:Ohms}
\end{figure*}

\begin{figure}[h]
	\centering
	\includegraphics[width=8cm]{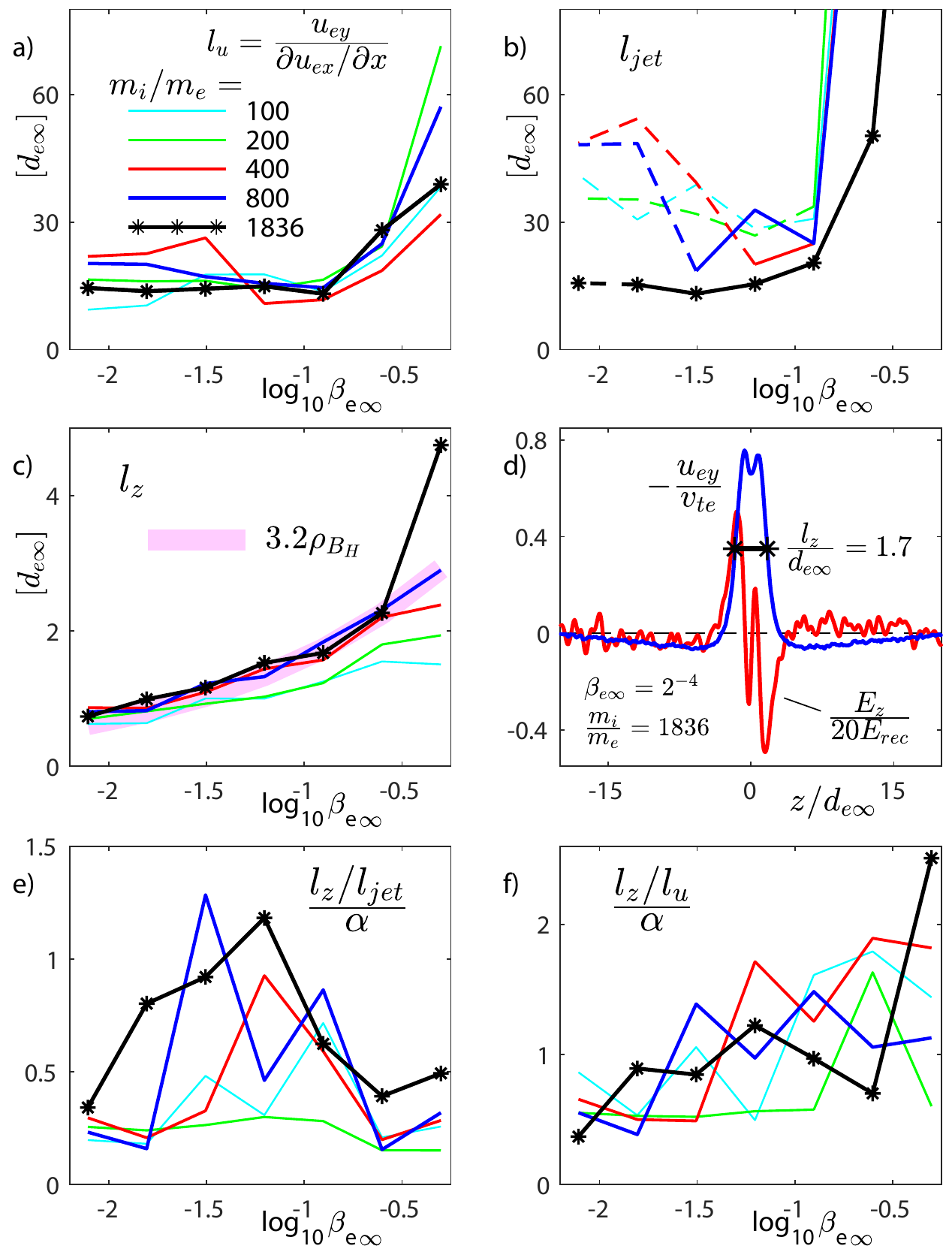}%
	\caption{a,b) Recorded values of $l_u$ and $l\scr{jet}$, which are similar to the inner and outer lengths of the EDR introduced in Refs.~\cite{karimabadi:2007,shay:2007}. In b) the dashed lines indicate runs in the non-adiabatic regime with enhanced $\pz$-heating observed for  $\beta_{e\infty}^2 m_i/m_e \lesssim  1$. Evidently, this regime is also characterized by an enhancement in the length of the EDRs (while $l_u$ remains near constant). c) Half-width of the electron current layer, which approximately scales as 3.2 $\rho_{B_H}$. Here $\rho_{B_H}$ is the typical electron Larmor radius at the upstream edge of the EDR. d) Considering the run with $m_i/m_e=1836$ and $\beta_{e\infty}=2^{-4}$, the blue line is $-u_{ey}/v_{te}$, while the red line show the profile of $E_z$. The black line (full-width) indicates how $l_z$= half-width-half-maximum is obtained form $-u_{ey}/v_{te}$.  e,f) Characteristic aspect ratios of the EDR, $l_z/l_{jet}$ and $l_z/l_u$, compared to the normalized reconnection rate $\alpha= E\scr{rec}/(v_A B)$.}  
	\hblabel{fig:Plots3L}
\end{figure}

Fig.~\ref{fig:Ohms} provides zoomed-in views of the EDRs recorded in the full set of runs applied in our study. As is typical for kinetic simulations of reconnection based on the Harris sheet geometry, after fast reconnection has commenced the reconnection rate initially increases and then reduces to a near steady state. Each of the profiles considered here correspond to times $t\simeq  70 \Omega_{ci}^{-1}$ during these later intervals of mostly steady state reconnection geometries. For $\beta_{e\infty} \geq 2^{-2} $ elongated EDRs  are mostly observed  where $l\scr{jet} > 100 d_{e\infty}$.  Here  the out-of-plane electron flows are largely provided by the diamagnetic drifts similar to those of the Harris sheet. Meanwhile, for $\beta_{e\infty} \leq 2^{-3}$, the force balance condition of the EDR as expressed in Eq.~\eq{BH} requires current densities boosted beyond those provided by the diamagnetic effect, and the EDRs acquire a shorter length. The marked ends of the jets are defined by the locations where $|u_{ex}|$ has fallen to 60\% of its respective peak value. 
For the adiabatic regime ($\beta_{e\infty}^2 m_i/m_e \gtrsim 1$), the lengths of the electron diffusion region are consistently observed to be $l\scr{jet}\simeq 20 d_{e\infty}$. For the regime $\beta_{e\infty}^2 m_i/m_e \lesssim 1$ of enhanced $\pz$ energization 
  \cite{egedal:2015}, a sharp transition occurs where the EDRs are characterized by  longer outflow jets.

In Fig.~\ref{fig:Plots3L} we show the values of 
 $l_u$ and $l\scr{jet}$ for all the runs applied in the study. The runs  with $\beta_{e\infty}\leq 2^{-3}$ (most relevant to the analysis in Section IV) are all characterized by $l_u\simeq 18 d_{e\infty}$. In Fig.~\ref{fig:Plots3L}(b) the dashed lines represent runs in the aforementioned regime $\beta_{e\infty}^2 m_i/m_e \lesssim 1$ of enhanced $\pz$ heating, and it is evident that this regime is characterized by larger values of $l\scr{jet}$.

\subsection{How continuity of electrons in the EDR sets the $l_u$ length scale}

Directly related to 
${\bf J}\scr{extra}$, we find that $l_u$ is shortened by the  enhancement of the electron flow into the EDR. 
Let $u_{ey0}$ be the flow in the $y$-direction at the $X$-line; given the definition of $l_u$ the flow in the $x$-direction is then $u_{ex}=u_{ey0} x/l_u$. For a uniform inflow velocity $u_{ez}$ mass continuity requires that 
\[
n   u_{ez} x=   n u _{ex}  l_z =  n u_{ey0} x  l_z /l_u
\]
where $l_z$ is the half width of the EDR analysed below. Carrying on, it follows that  
\[
l_u= \fd{u_{ey0} l_z}{u_{ez}} = \fd{B_H}{en\mu_0 u_{ez}}\quad.
\]
Meanwhile, the strong parallel streaming of electrons in toward the EDR causes the inflow speed of the electrons to increase. Empirically we find 
\[
u_{ez}\simeq 3.5 E\scr{rec}/B_{H}\quad.
\]
This enhancement of the inflow speed above $E\scr{rec}/B_{H}$ is caused by the inflow pressure anisotropy through the term
${{\bf J}_{\perp}}\scr{extra}$, and is consistent with the profiles in Figs.~\ref{fig:Fig1sup}(a,b).

As discussed above, the reconnection electric field, $E\scr{rec}$ adhere to the scaling laws of Ref.~\cite{shay:2004} where
\[ 
E\scr{rec}\simeq 0.1 v_A B_{1di} = \fd{0.1 B_{1di}^2}{\sqrt{\mu_0 n m_i}}\quad.
\]
It then follows that
\[
l_u \simeq \fd{B_H^2}{0.35B_{1di}^2} \sqrt{\fd{m_i}{m_e}} d_e\quad.
\]
Empirically and consistent with Fig.~\ref{fig:Fig1}(a), the magnetic field 1$d_i$ upstream of the EDR is approximately $B_{1di}\simeq B_{\infty}/2$. Furthermore, due to non-adiabatic effects $B_H$ depends on $m_i/m_e$ 
(see Fig.~\ref{fig:Plots2}(a)), and the scaling is consistent with $B_H^2\sqrt{m_i/m_e}/(0.35B_{1di}^2)\simeq 17$ for all the runs.  This scaling is also in agreement with recent results in Ref.~\cite{liu:2022}. In turn, this is in reasonable agreement with the observation that $l_u\simeq 18 d_{e\infty}$ in Fig.~\ref{fig:Plots3L}(a). 
It should be noted that $l_u$ varies within the duration of a single simulation, but  its precise length does  not  impact the qualitative predictions of the theory being presented below.

\subsection{The half-width, $l_z$, of the EDR}

The width of the EDR has previously been determined to scale with the orbit width of the meandering electrons \cite{horiuchi:1997,shay:1998, roytershteyn:2013}; that result is also consistent with the simulations presented here. In Fig.~\ref{fig:Plots3L}(c), the half-width $l_z$ is shown for each of the simulations. As illustrated in  Fig.~\ref{fig:Plots3L}(d), we define $l_z$ as the half-width where the half-maximum of the electron $u_{ey}$ flow velocity is recorded for $x=0$. 
We note that the simple approximation $l_z\simeq 2 d_{e\infty}$ represents a reasonable estimate; this is especially true for the normalized electron pressures most typical in the Earth's  magnetotail, $\beta_{e\infty} \simeq0.1$.

More accurately, the meandering orbit width of the electrons is related to the electron Larmor radius at the upstream edge of the EDR, $\rho_{B_H}=\sqrt{2m_e T_{e\perp}}/(eB)$, and can be estimated using  
$B=B_H$ in Eq.~\eq{BH}, and  $T_{e\perp} = T_{e\infty} B_H/B_{\infty}$. Thus, we then find
\begin{equation}
    \label{eq:rhoBH}
    \rho_{B_H}  = d_{e\infty} \left(\fd{B_{\infty}}{B_H} \beta_{e\infty}\right)^{\frac{1}{2}}
    = d_{e\infty} \left(\fd{12 }{\pi}\right)^{\frac{1}{8}} \beta_{e\infty}^{\frac{1}{4}}
    \quad.
\end{equation}
The thick magenta line in Fig.~\ref{fig:Plots3L}(c) represents $3.2 \rho_{B_H}$ and provides a good approximation for $l_z$ for the runs at $\beta_{e\infty}\leq 2^{-2}$ and $m_i/m_e\geq 400$. At the lowest values considered for $\beta_{e\infty}$, however, its is observed that $3.2 \rho_{B_H}$ underestimates $l_z$. This slightly weaker scaling of $l_z$ as a function of $\beta_{e\infty}$ may in part be explained by the fact that a fraction of the $u_{ey}$-drift is caused by the $E\times B$-drift of  the strong $E_z$ field at the upstream edge of the EDR. This field is illustrated by the red line in  Fig.~\ref{fig:Plots3L}(d) and the color profile in Fig.~\ref{fig:Dists}(a); for all the runs we find that the width of this $E_z$-structure is fixed when normalized by $d_{e\infty}$ (not shown).

Due to the strong drift provided by ${\bf J}\scr{extra}$ and because the currents inside the EDR are being driven by $\pz-\pp$, it is not mathmatically guaranteed  that the aspect ratio of the EDR is directly related to the normalized reconnection rate $\alpha$ discussed above and shown in Fig.~\ref{fig:Plots3mod10}(a). Therefore, in Figs.~\ref{fig:Plots3L}(e,f) we show the recorded values of $l_z/l_{jet}$ and $l_z/l_u$ both normalized by $\alpha$. In particular, it is observed that $(l_z/l_u)\simeq \alpha $ for most of the runs with $m_i/m_e \geq 400$. This quantity may then be useful for estimating the normalized reconnection rate from spacecraft observations \cite{burch:2020}.

\section{Breaking the frozen-in law at the reconnection $X$-line}

The major force terms of the EDR are ${\bf J}\times\B$ and  $\nabla\cdot {\bf p}_e$ which largely balance. As discussed above this force balance requirement can ultimately be expressed through the conditions in Eqs.~\eq{fire} or \eq{BH}. However, as seen in Fig.~\ref{fig:Fig1sup2mod}(a), in a small region ($4d_e\times 2d_e$) centered on the $X$-line the $({\bf J}\times\B)_y$ force vanishes, such that the more detailed force balance right at the $X$-line then requires that the RHS of Eq.~\eq{OhmsLawMod2} also  vanishes  (consistent with Fig.~\ref{fig:Fig1sup2mod}(b)). Thus, for the considered anti-parallel and symmetric configurations and within this limited region around the $X$-line,  the force balance constraint of Eq.~\eq{Ohmslaw} only involves the off-diagonal stress in the electron pressure tensor \cite{cai:1994}:
\begin{equation}
\hblabel{eq:Erec}
E\scr{rec}
= -\frac{1}{en}\left(\dd{p_{exy}}{x} + \dd{p_{eyz}}{z}\right)\quad.
\end{equation}
Again, here $E\scr{rec}$ is the $E_y$ electric field along the $X$-line (that is aligned with the $y$ axis and runs through $(x,z)=(0,0)$). By Faraday's law $E\scr{rec}$ then represents the rate at which magnetic flux crosses the $X$-line. Below we will derive a new model for the terms in Eq.~\eq{Erec}, providing a theory consistent with fast reconnection controlled by larger scale dynamics external to the EDR. 



%

\subsection{Meandering orbit motion of electrons within the EDR, setting the striated structure of the electron distribution function}

To elucidate how the upstream electron pressure anisotropy drives the current of the EDR, 
in Fig.~\ref{fig:Dists} we consider the trajectories of electrons with field-aligned velocities   ${\bf v}= 3.3v_{te} {\bf B}/B$ injected with $z=-4d_{e\infty}$  at various values of $x <0 $  just upstream of the EDR.  Here $v_{te}=\sqrt{T_e/m_e}$ is the electron thermal speed.  
The trajectories, computed through integration of the equations of motion with a Matlab ODE solver in the simulation fields, are representative because the upstream distributions (like in Figs.~\ref{fig:Distmime400k5}(a,d)) with  $\pze\gg\ppe$ ``feed'' electrons to the EDR with $|\vz|\gg\vp$ \cite{ng:2011}. In Figs.~\ref{fig:Dists}(a,b) the parts of the trajectories with $z<0$ ($z>0$) are represented by the blue (red) full lines.  To better illustrate the orbit morphology, for each transit across $z=0$ in Fig.~\ref{fig:Dists}(b) additional  orbits are initialized and shown by the dashed lines. This visualizes how the average force of $-e\vv\times B_z {\bf e}_z$ causes the meandering Speiser-type  motion \cite{speiser:1965} to diverge away from the $X$-line.

The EDR current is caused in part by the $E\times B$-drift of the   $E_z$ field displayed in Fig.~\ref{fig:Dists}(a). The large values  of $E_z$ are consistent with momentum balance with the strong gradients of $p_{\|}$ at the interface of the EDR. 
The initial blue section of the trajectories in Fig.~\ref{fig:Dists}(b)  are aligned with the magnetic field at the upstream edge of the EDR. Thus, the direction of the parallel thermal streaming is sensitive to the $B_y$ Hall magnetic field, which thereby sets the angle in the $xy$-plane at which the electrons are injected into the EDR region.

In Figs.~\ref{fig:Dists}(c-e) the colored and encircled triangles represent the particle velocities corresponding to the similarly colored velocity vectors in Fig.~\ref{fig:Dists}(b). For example, the magenta vector at $x=0$ in Fig.~\ref{fig:Dists}(b) pointing mostly in the $x$-direction, yields the velocity of an electron reaching the $X$-line directly from the upstream region, and the corresponding encircled magenta triangle in Fig.~\ref{fig:Dists}(d) has $|v_x|\gg |v_y|$. The $X$-line may also be reached through multiple meandering motions within the layer, steadily directing the velocity into the $-y$-direction. As shown in Figs.~\ref{fig:Dists}(c-e), repeating this procedure for four more initial velocities ${\bf v}= \vz {\bf B}/B$  within the interval  $0.1\leq \vz/v_{te}\leq  3.3$ yields the lines of colored triangles, corresponding to the centers of the striated structures of the EDR electron distributions  first described in Ref.~\cite{ng:2011}. In Figs.~\ref{fig:Dists}(c-e), the similar points marked by green circles are the result of injecting electrons along field lines from the opposing inflows of the EDR. Also, going from  $x/d_{e\infty}=-3$ to $x/d_{e\infty}=3$, we note how the displayed structures rotates in the $v_xv_y$-plane (and similarly in Fig.~\ref{fig:Distmime400k5}), corresponding to the rotation of the EDR current layer  direction mentioned above \cite{ng:2011,shuster:2015}, and parameterized by $l_u$.

\begin{figure}[h]
	\centering
	\hspace{-0.6cm}
	\includegraphics[width=9.1 cm]{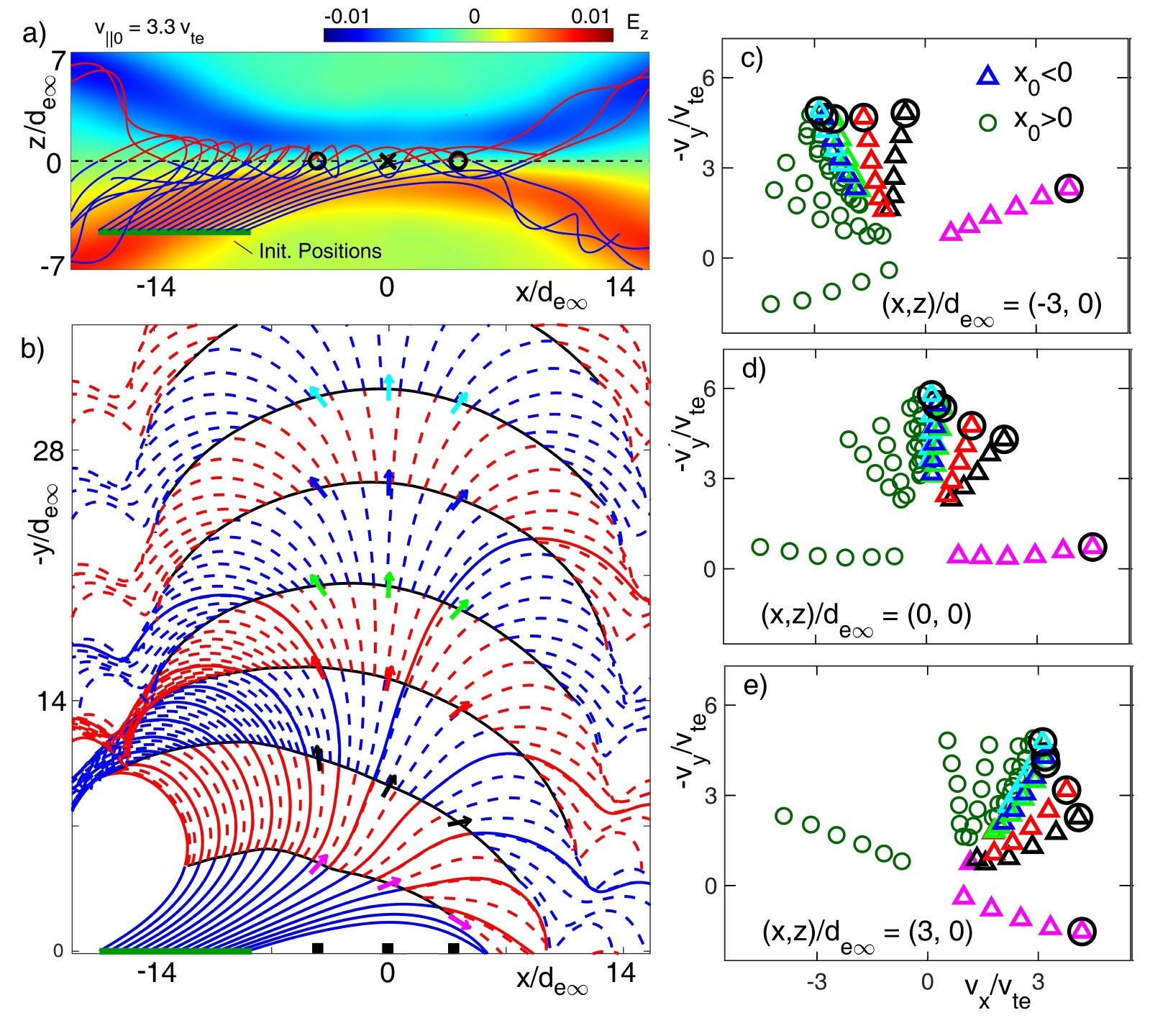}%
	\caption{a,b) Trajectories of  electrons injected along field lines into the EDR with $\vz=3.3v_{te}$. Three points, $(x,z/d_{e\infty})=(-3,0); (0,0); (3,0)$, are selected in a) indicating the spatial locations where the velocity arrows in b) are evaluated. As indicated, the panels c-e)  also correspond to these locations. 
	The angles indicated by arrows  in b) at which the trajectories reach the selected points, determine the location of the encircled triangles in c), d), and e), respectively. Similar trajectories are obtained by using a range of initial $\vz$ values and result in the striated structures marked by the symbols in c), characteristic of the EDR distributions \cite{ng:2011} and observed in Fig.~\ref{fig:Distmime400k5}.
	}
	\hblabel{fig:Dists}
\end{figure}

\begin{figure}[h]
	\centering
	\hspace{-0.6cm}
	\includegraphics[width=8.7 cm]{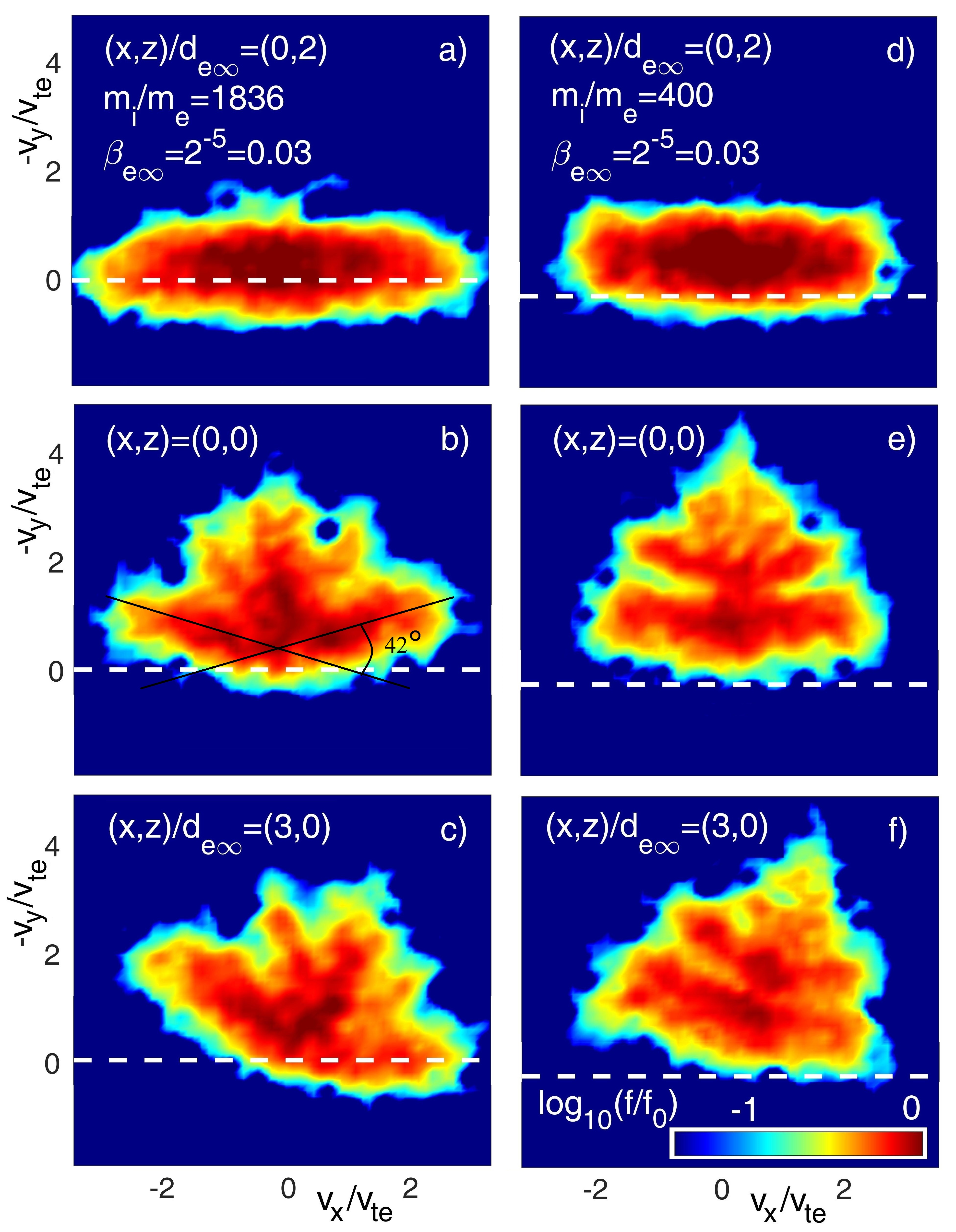}%
	\caption{Projections, $f_{xy}(v_x,v_y)= \int f({\bf v})dv_z$ of EDR electron distributions, recorded in simulations for $m_i/m_e=1836$ in a-c) and $m_i/m_e=400$ in d-f). The differences between the profiles are consistent with enhanced $\Delta p_{eyy,E}$ for $m_i/m_e=400$, caused by $\hat{E}\scr{rec,400}\simeq 2\hat{E}\scr{rec,1836}$.
	}
	\hblabel{fig:Distmime400k5}
\end{figure}

\subsection{Origin for the off-diagonal  $\ddt{p_{exy}}{x}$-stress at the reconnection $X$-line}

To understand how the upstream pressure anisotropy influences $p_{exy}$ we consider the electron distributions in Fig.~\ref{fig:Distmime400k5}. 
At the upstream edge of the EDR we have $p_{eyy}\simeq p_{ezz}$, and given Eq.~\eq{delp} it follows that  $p_{eyy}-p_{exx}=-\dpzp$. 
Comparing the distributions in Figs.~\ref{fig:Distmime400k5}(a,b)) it is clear that only a fraction of this upstream anisotropy is carried by the orbit motion to the $X$-line. Empirically at the $X$-line we find that  
\begin{equation}
\hblabel{eq:PeyyPexx}
\left.\left(p_{eyy}-p_{exx}\right)\right|\scr{X-line}= - \dfrac{4}{5} \dpzp+ \Delta p_{eyy,E} \quad.
\end{equation}
 The lines in  Fig.~\ref{fig:Distmime400k5}(b)) are drawn in ``by eye'' to indicate the center line of the striations in the $(v_x,v_y)$-plane, corresponding to  electrons reaching the X-line without bounces across $z=0$.
The factor of $4/5$ can be understood by noting the $\sim 42^\circ$ angle between these lines, which is double the angle at which the electrons are launched (by the ratio of $B_y/B_x$) as they initially enter the EDR.  The change in the straight distribution just upstream of the EDR (with  $\dpzp$) into the distribution with each side ``bent in the middle'' by $21^\circ$ then yields a simple estimate for the relative reduction in $p_{exx}-p_{eyy}$:
\[
\fd{\left.\left(p_{exx}-p_{eyy}\right)\right|\scr{X-line} } {\dpzp} = \cos^2(21^\circ) - \sin^2(21^\circ)/4 \simeq 0.84\quad. 
\]
The phase space density of the striations for the higher bounce numbers mainly add to $p_{eyy}$, explaining the further reduction in the ratio towards $4/5$. 

The additional term, $\Delta p_{eyy,E}$, in Eq.~\eq{PeyyPexx} corresponds to an increase in $p_{eyy}$ due to direct heating of electrons by the reconnection electric field. 
From Fig.~\ref{fig:Dists} it is clear that the ``tips'' of the triangular shaped distributions are composed of electrons meandering multiple times while streaming mostly in the $-y$-direction, energized most significantly by $\hat{E}\scr{rec}$. The  more pronounced triangular shapes  in Figs.~\ref{fig:Distmime400k5}(e,f) compared to Figs.~\ref{fig:Distmime400k5}(b,c) are consistent with stronger $\Delta p_{eyy,E}$ heating, as the value of $\hat{E}\scr{rec}$ for $m_i/m_e=400$ is about twice as large compared to that observed in the $m_i/m_e=1836$ run. 

As detailed with Figs.~\ref{fig:Dists} and \ref{fig:Distmime400k5}, within the EDR the $xy$-projections of the electron distributions rotate approximately as a ``solid-body'' \cite{ng:2011,divin:2016} such that for small $x/l_u$
\begin{equation}
\hblabel{eq:Pexy}
p_{exy} =\frac{x}{l_u}\left.\left(p_{eyy}-p_{exx}\right)\right|\scr{X-line}\quad.
\end{equation}
Combining Eqs.~\eq{PeyyPexx} and \eq{Pexy} we then obtain
\begin{equation}
\hblabel{eq:dPexy}
\dd{p_{exy}}{x} =\frac{1}{l_u}\left(-\dfrac{4}{5} \dpzp+\Delta p_{eyy,E}\right)\quad.
\end{equation}

\subsection{Origin for the off-diagonal  $\ddt{p_{eyz}}{z}$-stress at the reconnection $X$-line}

In the present paper we do not provide a formal derivation of the 
$\ddt{p_{eyz}}{z}$-stress.  However, the first term in Eq.~\eq{dPexy} is independent of $E\scr{rec}$ as it is imposed by the upstream pressure anisotropy. The motion of each electron in the X-line region is naturally governed by Newton's laws, and once all forces are summed up over the electrons in a fluid element, the terms independent of $E\scr{rec}$ must cancel in order for the fluid to be in force balance. Thus,  force balance at the X-line requires that the $\ddt{p_{exy}}{x}$ stress imposed externally is cancelled by an opposite term in
Eq.~\eq{Erec}. The only term available for this cancellation is  
$\ddt{p_{eyz}}{z}$, and this simple argument then suggests that
\begin{equation}
\hblabel{eq:dPeyz}
\dd{p_{eyz}}{z} =\frac{1}{l_u}\dfrac{4}{5} \dpzp \quad.
\end{equation}



Some theoretical support for Eq.~\eq{dPeyz} can be derived from the results in 
Refs.~\cite{kuznetsova1998kinetic,hesse:1999},  which also obtained  approximations for the off-diagonal pressure tensor derivatives ($\ddt{p_{exy}\sscr{Hesse}}{x}$ and $\ddt{p_{eyz}\sscr{Hesse}}{z}$). 
\[
\dd{p_{exy}\sscr{Hesse}}{x}=
\dd{p_{eyz}\sscr{Hesse}}{z}= n\sqrt{\fd{m_e T_e}{2}}\, \dd{u_{ex}}{x}\quad.
\]
These were derived through an assumption of isotropic upstream electron pressure. As such, given that 
upstream of the EDR $p_{exx}- p_{eyy} = \Delta p_{\|,\perp}$ is large, the  $\ddt{p_{exy}\sscr{Hesse}}{x}$-approximation is inaccurate and is not consistent with the simulation data presented below.
In contrast, the conditions required for the accuracy of $\ddt{p_{eyz}\sscr{Hesse}}{z}$ are well satisfied, as the upstream pressure in the $yz$-plane is isotropic, $p_{eyy}\simeq p_{ezz}\simeq \ppe$. 
Still, we find that a scale-factor must be applied, $\ddt{p_{eyz}}{z} \simeq 3.6\,\ddt{p_{eyz}\sscr{Hesse}}{z}$, for matching our numerical results at $m_i/m_e=1836$ and $\beta_{e\infty}\leq 2^{-4}$ (see below).

 Using the result above that the width of the EDR is $l_z\simeq3.2\rho_{B_H}$,  
from Amp\`{e}re's law it follows that $u_e= B_H/(\mu_0e n l_z)= B_H^2/(3.2\,\mu_0nm_e\vp)$. 
With  $\ddt{u_{ex}}{x}=u_e/l_u$, and following Refs.~\cite{kuznetsova1998kinetic,hesse:1999} 
we find
\begin{eqnarray}
\hblabel{eq:Phesse2}
\dd{p_{eyz}}{z}
&\simeq&
 3.6  \dd{p_{eyz}\sscr{Hesse}}{z} \simeq
\fd{3.6}{3.2} \sqrt{\frac{m_e T_{e\perp}}{2}} \frac{B_H^2}{\mu_0m_e\vp l_u}
\nonumber \\[2ex]
&\simeq&  \fd{9}{16} \frac{B_H^2}{\mu_0 l_u} \simeq0.804\frac{\dpzp}{l_u}
\simeq \frac{4}{5} \frac{\dpzp}{l_u}
\,\,.
\end{eqnarray}
As illustrated by the black contours in Figs.~\ref{fig:Fig1}(c), for the present configurations the momentum balance in Eq.~\eq{BH} corresponds to the condition $0.7 B_H^2/\mu_0\simeq \dpzp$. This substitution was applied in the last line of  Eq.~\eq{Phesse2}. 
Below we will show how the scale factor of 3.6 in front of $\ddt{p_{eyz}\sscr{Hesse}}{z}$ is supported by the numerical data obtained at $m_i/m_e=1836$, $\beta_{e\infty}\leq 2^{-4}$ (see Fig.~\ref{fig:Fig11}(a)), and its theoretical origin will be the subject of further investigations.

\subsection{The term breaking the electron frozen in law}

Given the described cancellation of terms, the theory suggests that the electron dynamics of the EDR do not represent a bottleneck for reconnection.
Again, inserting Eqs.~\eq{dPexy} and \eq{dPeyz} into Eq.~\eq{Erec} only one term remains for balancing the reconnection electric field
\begin{equation}
\hblabel{eq:Erec2}
E\scr{rec}\simeq -\frac{1}{enl_u} \Delta p_{eyy,E}\quad.
\end{equation}
Separate from the force balance constraint expressed in Eq.~\eq{Erec2} we can  obtain a second independent estimate for $\Delta p_{eyy,E}$: In Fig.~\ref{fig:Dists} we observe that electrons in the $X$-line region typically travel a distance between $0$ and 2$l_u$ in the $y$-direction such that $\Delta p_{eyy,E}\simeq -e n l_u E\scr{rec}$. Applying this in Eq.~\eq{Erec2} we obtain the seemingly mundane expression that
\begin{equation}
    E\scr{rec}\simeq E\scr{rec}\quad.
    \hblabel{eq:ErecErec}
\end{equation}
Nevertheless, this is an important result, as it shows that the present  geometry can accommodate any reconnection rate imposed at larger scales onto the EDR region. 
This is consistent with the near-uniform ``ion-normalized-reconnection'' rate, $\alpha$, in Fig.~\ref{fig:Plots3mod10}(a). It also confirms previous numerical results and theoretical conjectures that, with the formation of electron scale current layers, the electron dynamics can readily accommodate the rates of reconnection imposed by the plasma behavior at  larger scales \cite{shay:1999,stanier:2015,liu:2017}. 
We also note that because $E\scr{rec}\propto \Delta p_{eyy,E}/l_u$ and $\Delta p_{eyy,E}\propto l_u$, the reconnection rate becomes independent of $l_u$.  In Appendix C we include additional discussion of how our interpretation of Eqs.~\eq{Erec2} and \eq{ErecErec} are not in conflict with results of resistive fluid models.


\subsection{Numerical evidence for the validity of the model}

As described above, the theory is aided  by fully kinetic VPIC simulations  carried out for a matrix of  $\beta_{e\infty}$ and $m_i/m_e$ values.  We next validate each of  the separate theoretical predictions  against the numerical results. 
First, the described dynamics involving trapped electrons require that within the inflow regions $|u_{e\|}|\ll v_{te}$, which is valid for $\beta_{e\infty}\gtrsim \sqrt{m_e/m_i}$ \cite{egedal:2015,le:2015}, and the predicted profiles  $B_H$ are confirmed in  Figs.~\ref{fig:Plots2}(a). 
\begin{figure}[h]
	\centering
	\includegraphics[width=8.6 cm]{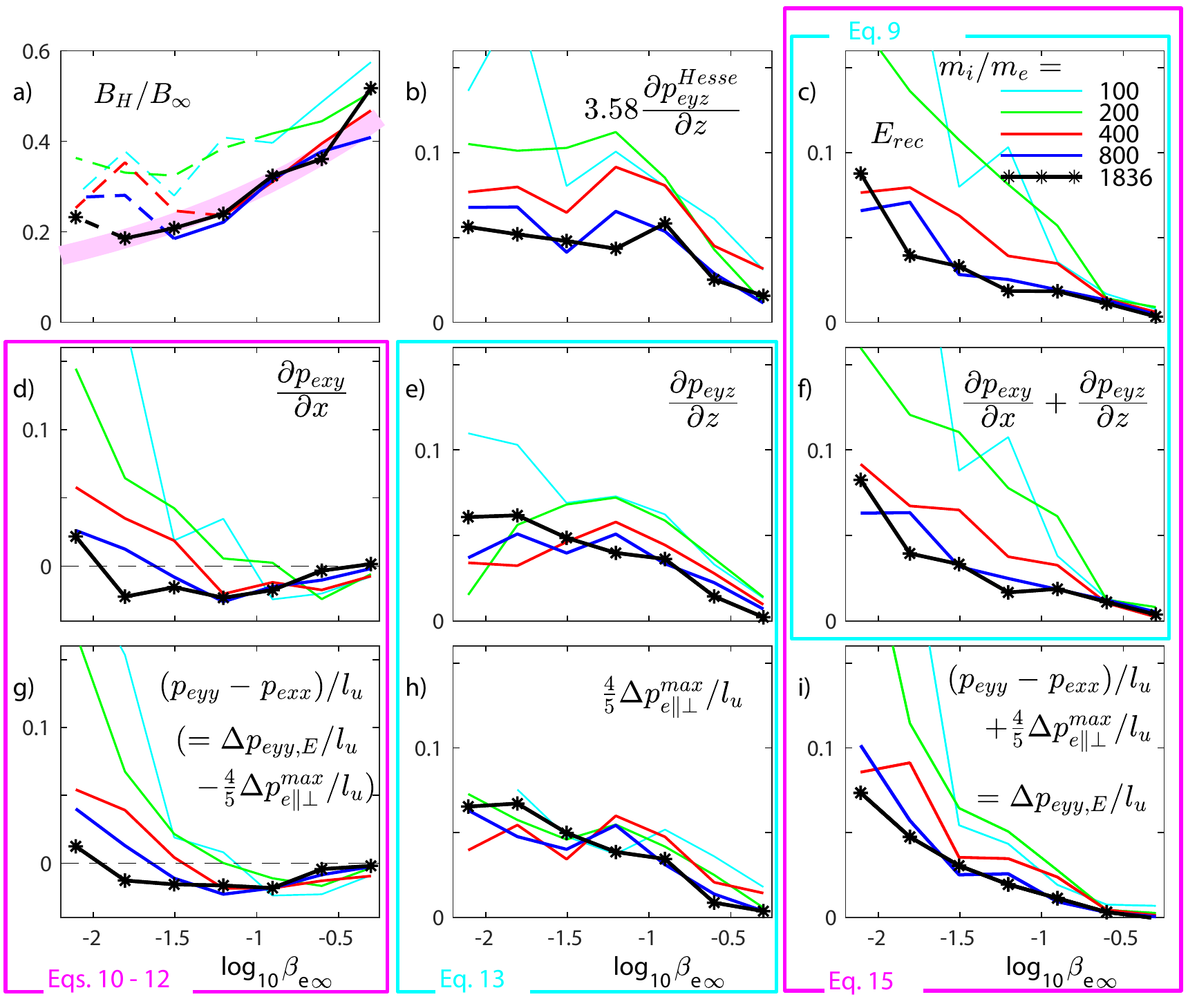}%
	\caption{Obtained from a matrix of 35 kinetic simulations, profiles as functions of $\log_{10}(\beta_{e\infty})$ are shown for a range of parameters as marked in the panels. In a) the magenta line is Eq.~\eq{BH}. In c) $E\scr{rec}$ is normalized by $T_{e\infty}/(ed_{e0})$ (which is different from $\hat{E}\scr{rec}$ in Fig.~\ref{fig:Plots3mod10}(b)). The profiles in b), d--i) are various electron pressure tensor derivatives, normalized by $-n_{\times}T_{e\infty}/(ed_{e0})$, such that the resulting dimensionless values can be compared directly to those of $E\scr{rec}$ in c). 
	}
	\hblabel{fig:Plots2}
\end{figure}

For all the numerical runs in our study, the individual contributions of $\ddt{p_{exy}}{x}$ and $\ddt{p_{eyz}}{z}$ are shown in Figs.~\ref{fig:Plots2}(d,e), respectively.
In Fig.~\ref{fig:Plots2},  ${E}\scr{rec}$ is normalized by $T_e/(ed_{e0})=0.024$, and for direct comparison, all terms  involving pressure derivatives are normalized by  $-n_{e\times} T_{e\infty}/d_{e0}$. With this normalization, confirming Eq.~\eq{Erec}, the sum of
$\ddt{p_{exy}}{x}+\ddt{p_{eyz}}{z}$ in Fig.~\ref{fig:Plots2}(f) reproduces with good accuracy $E\scr{rec}$ in Fig.~\ref{fig:Plots2}(c).

Eqs.~\eq{PeyyPexx}-\eq{dPexy} represent a key insight for the theory which is confirmed numerically, as Fig.~\ref{fig:Plots2}(g) provides an accurate representation of Fig.~\ref{fig:Plots2}(d). 
Consistent with recent spacecraft observations \cite{egedal:2019}, $\ddt{p_{exy}}{x}$ is mostly negative for runs with $m_i/m_e\geq 400$, and is thus dominated by the external stress  imposed by $\pze\gg\ppe$. Likewise, Eq.~\eq{dPeyz} is supported by the data, as the 
profiles of $\frac{4}{5}\dpzp/l_u$  in Fig.~\ref{fig:Plots2}(h), are observed to provide a match to $\ddt{p_{eyz}}{z}$ in Fig.~\ref{fig:Plots2}(e). This is especially the case for $m_i/m_e\geq 400$, corresponding to the adiabatic limit required for the validity of Eq.~\eq{BH}. Also in this limit, the sums shown in Fig.~\ref{fig:Plots2}(i) of the profiles of Figs.~\ref{fig:Plots2}(g,h)  provide a good match to $E\scr{rec}$ in Fig.~\ref{fig:Plots2}(c), confirming Eq.~\eq{Erec2}. 

\begin{figure}[h]
	\centering
	\includegraphics[width=0.47\textwidth]{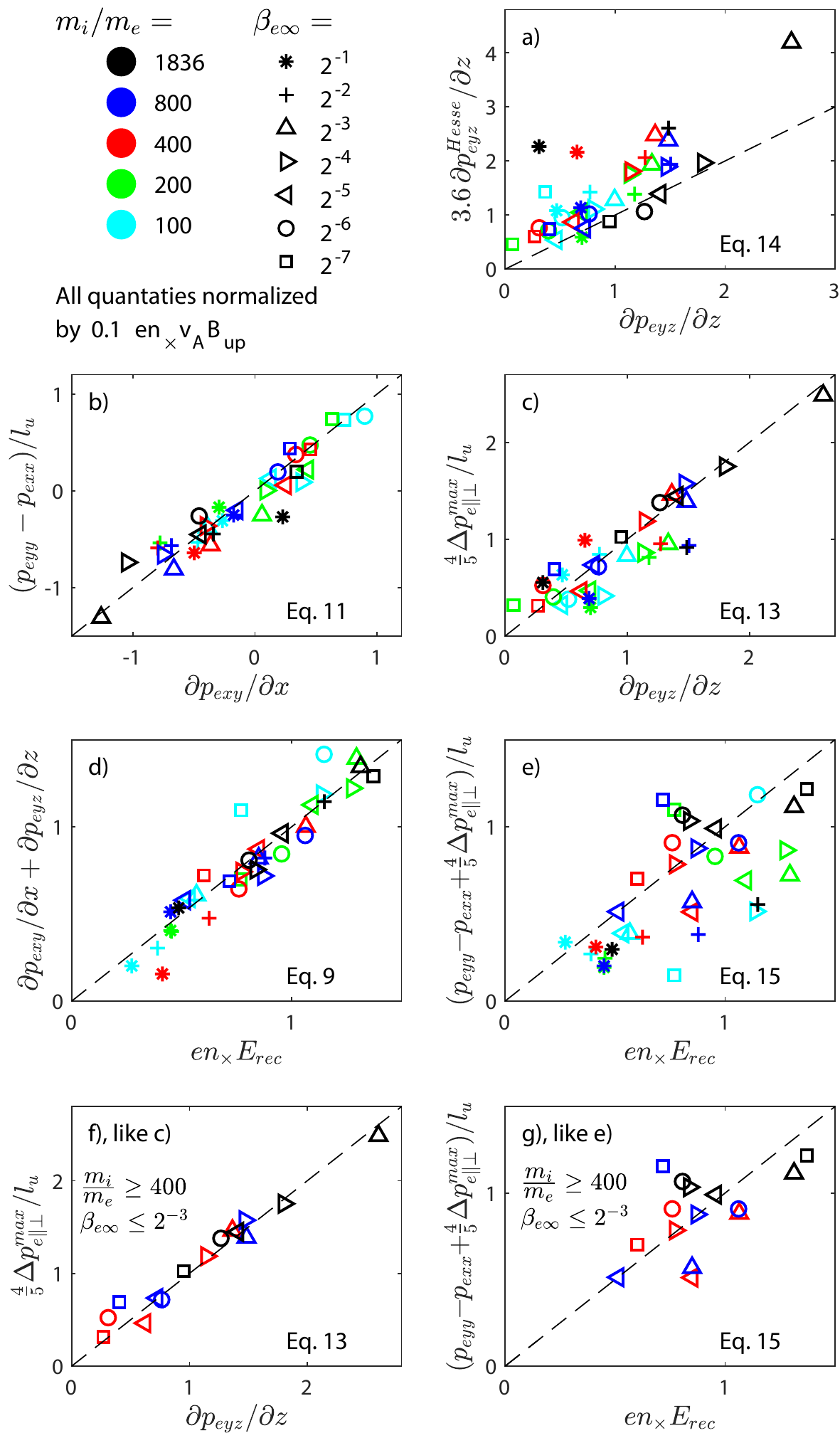}%
	\caption{ As indicated in the panels of a-g), the LHS and RHS of key equations of the manuscript are tested using data collected in the numerical runs.   
 Most significantly, in b) Eq.~\eq{Pexy} is confirmed, and in f) Eq.~\eq{dPeyz} is confirmed for the important parameter range  of $m_i/m_e \geq 400$ and $\beta_{e\infty} \leq 2^{-3}$.
 To ensure that data from the runs with the largest values of $\beta_{e\infty}$ do not cluster close to $(0,0)$,  all terms are normalized by $0.1e n_{e\times} v_A B_{up}$.
 }
	\hblabel{fig:Fig11}
\end{figure}

To further validate the most important equations in the study, in Fig.~\ref{fig:Fig11} we plot the terms of Eqs.~\eq{Erec}, \eq{Pexy}, \eq{dPeyz}, \eq{Phesse2}, and \eq{Erec2}, with all the shown quantities normalized by $0.1 e n_{e\times} v_A B_{up}$. First, in panel a) the factor of $\sim3.6$ used in Eq.~\eq{Phesse2} is consistent with the data at $m_i/m_e=1836$, $\beta_{e\infty}\leq 2^{-4}$. In panels b) and c) the approximations introduced for $\ddt{p_{exy}}{x}$ and $\ddt{p_{eyz}}{z}$ are validated. For the case of 
$\ddt{p_{eyz}}{z}$ panel f) indicates an improved accuracy of the approximation for the most relevant parameters of $m_i/m_e \geq 400$ and $\beta_{e\infty}\leq 2^{-3}$. From panel d) it is clear that the momentum balance Eq.~\eq{Erec} is well satisfied by the simulations. Panel e) tests the penultimate  expression (Eq.~\eq{Erec2}) of the paper. Because of the often opposing signs of the $\ddt{p_{exy}}{x}$ and $\ddt{p_{eyz}}{z}$ estimates, when calculating their sum the relative uncertainty/error is enhanced, explaining the enhanced scatter of the data points. However, as shown in panel g), when limiting the simulations to those with 
 $m_i/m_e \geq 400$ and $\beta_{e\infty}\leq 2^{-3}$ most predictions for $E\scr{rec}$ fall within 20\% of the correct value and all predictions are within 50\%.

\section{Summary and discussion}

The use of the VPIC code implemented in a modern super-computing facility enabled a matrix of simulations to be performed for anti-parallel magnetic reconnection. This matrix   spans a range of the normalized electron pressures, $\beta_{e\infty}$, as well as the ion to electron mass ratios, $m_i/m_e$.  Our study reveals a range of results, which require new interpretations of the electron dynamics of the EDR and inner IDR that are different from previous models including those in Refs.~\cite{hesse:1999,drake:2008,daughton:2006}. 
A main difference from the numerical studies a decade or more ago is  our recent  ability to carry out routine kinetic simulations at the natural proton to electron mass ratio of  $m_i/m_e=1836$. At full mass ratio the effect of the electron pressure anisotropy that develops in the reconnection inflow becomes the dominant force-term not only within the EDR but also for the inner part of the  IDR.

It has previously been determined \cite{le:2009,egedal:2013} that upstream of the EDR strong electron pressure anisotropy with $p_{e\|} \gg p_{e\perp}$ is driven by the convection of electrons into  the reconnection region characterized by low values of $B$. This upstream pressure anisotropy is responsible for driving the strong electron currents within the EDR. In addition, with Fig.~\ref{fig:Fig1sup2mod} we visualize  how the electron inflow speed for locations inside the IDR (but outside the traditional EDR) exceeds the inflow speed of the magnetic field by a factor up to 3.5. This enhances the Hall magnetic field perturbation beyond what can be expected from traditional ``whistler reconnection'' \cite{drake:2008}. 
Nevertheless, it is still possible that  the condition $({\bf J}\times \B)_y\simeq (\nabla \cdot {\bf p}_e)_y$ is compatible with dispersive waves \cite{rogers:2001,cassak:2015}.
The strong Hall field $B_y$ perturbations at the edge of the EDR are mainly a consequence of dynamics related to the electron pressure anisotropy yielding a new and dominate force balance constraint, namely $({\bf J}\times \B)_y\simeq (\nabla \cdot {\bf p}_e)_y \gg en_eE\scr{rec}$. Furthermore, the enhanced perpendicular flow into the EDR by ${\bf J}\scr{extra} = [(\pz-\pp)/B]{\bf b} \times \mathbf{\kappa}$  must be sourced by electrons flowing along the field lines on the inflow-side of the separatrix layers. This is consistent with spacecraft observations that the Hall currents extend many tens of $d_i$ away from the EDR \cite{manapat:2006}.

Aside from the results summarized above, the main goal of this paper is to develop a theory that can account for the electron momentum balance directly at the $X$-line within the center of the EDR for anti-parallel reconnection. Previously, Refs.~\cite{kuznetsova1998kinetic,hesse:1999} provided approximate expressions for the off-diagonal stress terms $\ddt{p_{exy}}{x}$ and
$\ddt{p_{eyz}}{z}$. After applying a scale factor, their theoretical form for $\ddt{p_{eyz}}{z}$ is in agreement with our numerical results. Meanwhile, the electron pressure anisotropy that develops upstream of the EDR turns out to have a large impact on  $\ddt{p_{exy}}{x}$ not previously considered. Our updated theory for
$\ddt{p_{exy}}{x}$ is cast in two separate terms. The first term is caused by the upstream electron pressure anisotropy and cancels the stress of $\ddt{p_{eyz}}{z}$. In hindsight, this cancellation is not surprising as the EDR current, mainly driven by the upstream $p_{e\|}\gg p_{e\perp}$, must be in force balance independent of the value of $E\scr{rec}$. 

The additional second term for  
 $\ddt{p_{exy}}{x}$ in Eq.~\eq{dPexy} is the term that scales with $E\scr{rec}$ and as such, can be considered the term that actually breaks the electron frozen-in condition at the $X$-line. This term is related to the  increase of the $p_{eyy}$ pressure tensor element caused directly through heating by $E\scr{rec}$. Because the electron flow in the $y$-direction is largely  fixed by the upstream $p_{e\|} \gg  p_{e\perp}$ (see Eq.~\eq{BH}), the increase in $p_{eyy}$ becomes linear in $E\scr{rec}$, and the rotation with length scale $l_u$ of $p_{eyy}$ in the $xy$-plane then yields the stress required to balance $E\scr{rec}$. Because the term is proportional to $E\scr{rec}$, it facilitates reconnection at any external rate imposed onto the EDR. 
From the perspective of the electrons the reconnection rate is $\hat{E}\scr{rec}$ shown in Fig.~\ref{fig:Plots2}(c), and runs with low values of  $m_i/m_e$ have significant enhanced values of  $\hat{E}\scr{rec}$. 
Even so, the identified reconnection mechanism can still accommodate these artificially enhanced values of $\hat{E}\scr{rec}$. 
 
 In closing, we note that our new theory predicts that $\ddt{p_{exy}}{x}$ and $\ddt{p_{eyz}}{z}$ have opposite signs for $m_i/m_e=1836$. This result has been directly confirmed in spacecraft observations of an EDR encounter by MMS in the Earth's magnetotail \cite{egedal:2019}. Other MMS observations also support the approximation that reconnection in the Earth's magnetotail occurs in regimes consistent with laminar 2D kinetic models. One reason for the apparent success of such laminar 2D models is perhaps that for $m_i/m_e=1836$ the force by $E\scr{rec}$ is  small compared to the forces associated with electron pressure anisotropy, such that $E\scr{rec}$ only slightly perturbs the electron orbit dynamics within the EDR. Again, from the perspective of the electrons the reconnection rates imposed by the inertia and dynamics of the much heavier ions are in fact very modest and the small modifications imposed by $E\scr{rec}$ on the electron motion is not sufficient to drive strong instabilities. In order for instabilities to alter the momentum equation in Eq.~\eq{Ohmslaw} they much have an inverse growth rate similar to the short electron transit time through the EDR. For example, the Lower-Hybrid-Drift-Instability (LHDI) may perturb the out-of-plane structure of the EDR, but this does not necessarily cause a fundamental change in its underlying 2D dynamics \cite{le:2017,greess:2021,schroeder:2022}. Given recent progress in the understanding of how the larger scale ion dynamics influence the reconnection rate \cite{stanier:2015,liu:2017}, a more detailed and complete picture for reconnection now emerges consistent with local observations of reconnection within the Earth's magnetotail.


\section*{DATA AVAILABILITY}
The data that support the findings of this study are available
from the corresponding author upon reasonable request.  In addition, using the initial conditions specified in the text, the data can be reproduced with the open source VPIC
code available online (https://github.com/lanl/vpic) (https://zenodo.org/record/4041845\#.X2kA1x17kuY;
https://doi.org/10.5281/zenodo.4041845).

\section*{Appendix A: Basic mechanisms governing the formation of electron pressure anisotropy upstream of the EDR}

As described in Sec.~II for anti-parallel reconnection strong electron pressure anisotropy with $\pz\gg\pp$ develops within the inflow regions. While this has been the topic of previous investigations \cite{egedal:2008jgr,le:2009,egedal:2013}, for the convenience of the reader we will here provide a short description of the physical mechanisms that governs the generation of the anisotropy. Within the inflow regions the electrons  follow the magnetic flux-tubes in their convection toward the EDR, while the ions are  unmagnetized.  By their inertia, the ions  decouple from the motion of the magnetic field, and they dictate a near-uniform plasma density within the region. In turn, the electrons respond strongly to match this uniform density and maintain quasi-neutrality.  
As illustrated in   Fig.~\ref{fig:Fig1}(a),  the declining magnetic field strength $|\B|$  causes the widths of the magnetic flux-tubes to expand
as the EDR is approached. To avoid a reduction in the  electron density, field-aligned electric fields $E_\|$ develop \cite{egedal:2009pop}, compressing  the range of the parallel motion for trapped electrons. This boosts the electron density such that quasi-neutrality ({\sl i.e.}~$n_e\simeq n_i$) is maintained. The profiles of  $E_\|$ in many cases trap  all thermal electrons, limiting thermal heat conduction  and yielding a regime that differs significantly from the standard Boltzmann regime where $T_e$ is constant. 
 
For the trapped electron population, we  note that the area of a given flux-tube scales as $1/B$, and the total number of trapped particles in the flux-tube section (of length $l$) therefore scales as $N\scr{trapped}\propto n\scr{trapped} l /B$. For the case where the trapped electrons dominate the full  distribution, particle conservation then requires that $l\propto B/n\scr{trapped}$. Next, similar to Fermi heating, the conservation of the parallel action for each trapped electron requires that $l\vz$ is constant such that $\vz\propto n\scr{trapped}/B$, yielding  $T_{e\|}\propto \vz^2\propto n\scr{trapped}^2/B^2$. Furthermore, because $\mu=m\vp^2/(2B)$ is conserved it is  clear  that $\vp^2 \propto B$, such that for this  trapped electron population $T_{e\perp}\propto B$. It then follows that $\pz=nT_{e\|}\propto n\scr{trapped}^3/B^2$ and $\pp=nT_{e\perp}\propto n\scr{trapped}B$, coinciding with the CGL-scaling laws \cite{chew:1956}. Again, more accurate scaling laws are  provided in Refs.\cite{le:2009,egedal:2013}, taking into account that not all electrons become trapped. This yields a smooth transition from Boltzmann scaling ($\pz=\pp=nT_e$) at low values of $n/B$ to the CGL scalings at large values of $n/B$.

 \section*{Appendix B: Force balance along  the EDR }
\begin{figure}[h]
	\centering
	\includegraphics[width=8.5cm]{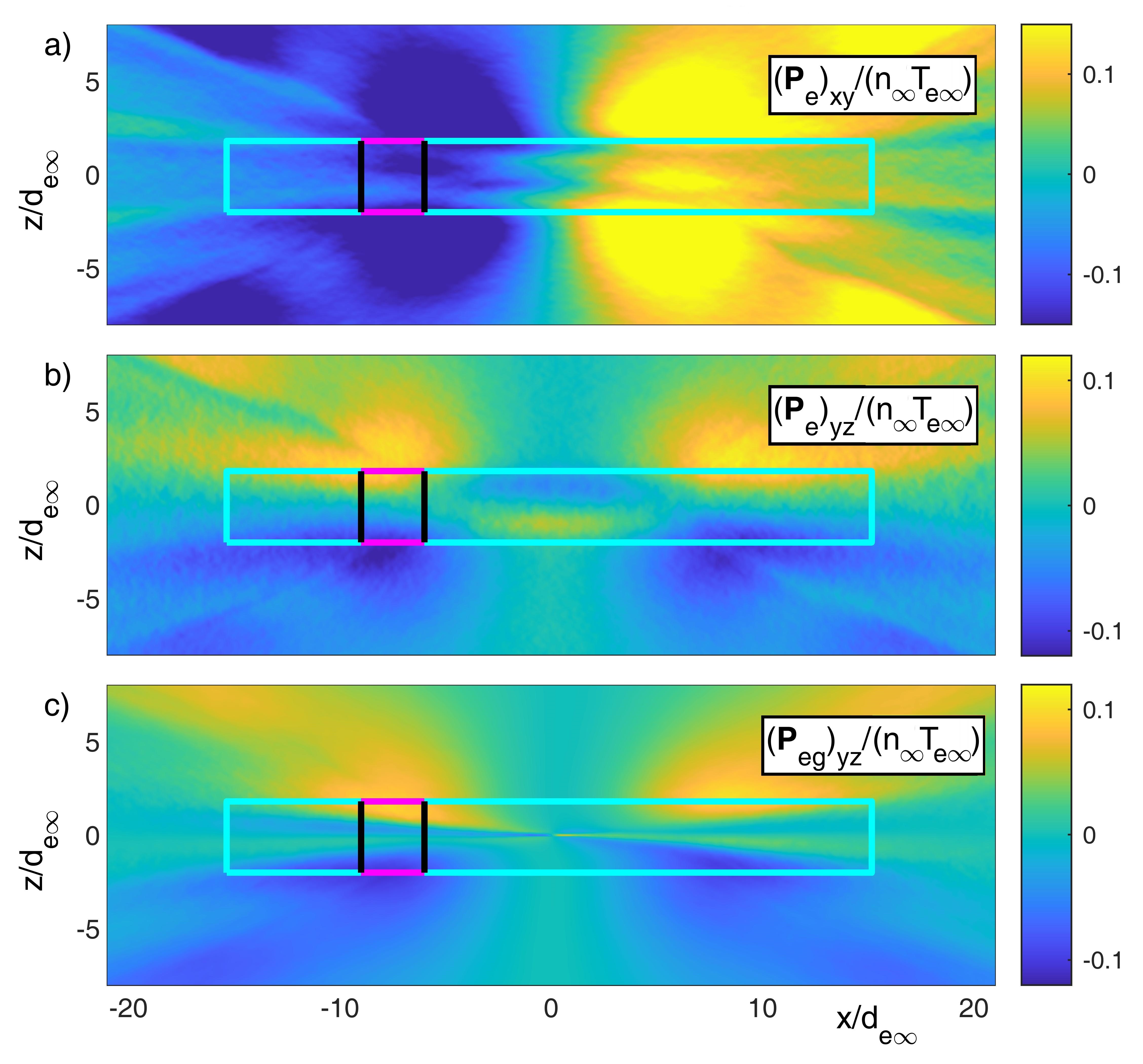}%
	\caption{Considering the same simulation ($m_i/m_e=1836$ and $\beta_{e\infty}=2^{-4}$) as in Fig.~\ref{fig:Fig1}, color contours are presented for a) ${p_{e}}_{xy}$, b) ${p_e}_{yz}$, and c) ${p_{eg}}_{yz}$. The EDR is outlined by the cyan rectangle, and a fluid-element is selected for analysis by the black/magenta square.   }
	\hblabel{fig:Fig1sup}
\end{figure}

 When away from the vicinity of the $X$-line, the $J\times B$-forces within the EDR are large (see Fig.~\ref{fig:Fig1sup2mod}(a)), and  in Fig.~\ref{fig:Fig1sup} we explore how
the large values of $({\bf J}\times\B)_y$ are related to the electron pressure anisotropy that forms upstream of the EDR. For this, consider the small fluid element which in each panel is marked  by the black/magenta square. Naturally, we can integrate Eq.~\eq{OhmsLawMod2} over the surface of this fluid element. For the RHS of Eq.~\eq{OhmsLawMod2} the largest contribution is from $\int (\nabla\cdot {\bf p}_e)_y dA$.
In turn, by the divergence theorem
$\int (\nabla\cdot {\bf p}_e)_y dA = \Delta {\bf p}_{exy}\, \Delta z + \Delta {\bf p}_{eyz}\, \Delta x$.
Here, $\Delta {\bf p}_{exy}$ is the change in $ {\bf p}_{exy}$ between the two black sides of the element, while $\Delta {\bf p}_{eyz}$ is the change in  
${\bf p}_{eyz}$ between the two magenta sides. Comparing Figs.~\ref{fig:Fig1sup}(a,b) it is clear that $|\Delta {\bf p}_{exy}|\ll|\Delta {\bf p}_{eyz}|$, such that 
$\int (\nabla\cdot {\bf p}_e)_y dA \simeq \Delta {\bf p}_{eyz}\, \Delta x $.

We have now demonstrated that $\int ({\bf J}\times\B)_y dA \simeq \Delta {\bf p}_{eyz}\, \Delta x $. On the other hand, ${\bf J}\times\B= \nabla\cdot {\bf T}$, where 
$ {\bf T} = (B^2/\mu_0) ({\bf bb}-{\bf I}/2)$ is the Maxwell stress tensor. By similar arguments to those applied when integrating $\int (\nabla\cdot {\bf p}_e)_y dA$, we then find that $\int ({\bf J}\times\B)_y dA = 
\int (\nabla\cdot {\bf T})_y dA \simeq  \Delta {\bf T}_{yz}\, \Delta x$.  Note that for the fluid element with $\Delta x\simeq \Delta z$, we have $\Delta {\bf T}_{yz}\, \Delta x \gg \Delta {\bf T}_{xy}\, \Delta z$ due to the $\simeq 1/10$ aspect ratio of the full EDR. 
Neglecting $E_y$, the force balance constraint in  Eq.~\eq{OhmsLawMod1} can then be expressed as 
$\Delta {\bf T}_{yz} \simeq \Delta {\bf p}_{eyz}$, and given these elements are asymmetric about $z=0$ this further reduces to ${\bf T}_{yz} \simeq {\bf p}_{eyz}$ along the upstream edge of the EDR. 
An identical  analysis can be carried out for the forces in the $x$-direction, which yields the similar result that 
${\bf T}_{xz} \simeq {\bf p}_{exz}$ along the upstream edge of the EDR. 

The integrals over the fluid element involved the full pressure tensor ${\bf p}_e$, but by the divergence theorem the results only depend on the values of  ${\bf p}_e$ at the upstream edge of the EDR; here ${\bf p}_e$ is well approximated by its CGL-form  ${\bf p}_e\simeq {\bf p}_{eg} = (\pz-p_{e\perp}) {\bf bb}+ p_{e\perp} {\bf I}$. To illustrate that ${\bf p}_e\simeq {\bf p}_{eg}$ outside the EDR, in 
Figs.~\ref{fig:Fig1sup}(b,c) the profiles of  ${\bf p}_{eyz}$ and ${\bf p}_{eg,yz}$ can be compared directly. Simple manipulations then show that the two force balance constraints that ${\bf T}_{yz} \simeq {\bf p}_{eyz}$ and 
${\bf T}_{xz} \simeq {\bf p}_{exz}$, can now be expressed on a common form through the marginal firehose condition $\pz-p_{e\perp}= B^2/\mu_0$, again, applicable along the upstream edge of the EDR.  

To summarize this analysis, we have shown that for the present simulation at $m_i/m_e=1836$ that 
$|en_eE_y| \ll |({\bf J}\times\B)_y|\simeq |(\nabla\cdot {\bf p}_e)_{y}|$. By integrating over a fluid element that spans the width of the EDR, it is then found that the strong ${\bf J}\times \B$-forces  within the EDR are balanced by the force of the pressure anisotropy at the upstream edge of the EDR. In turn, at the upstream edge of the EDR the L\^{e}-2009 equations of state \cite{le:2009} are applicable, and together with the force balance condition $\pz-p_{e\perp}= B^2/\mu_0$, this leads to the scaling law for the current across the EDR as expressed in Eq.~\eq{BH}. It is noteworthy here how the current across the EDR is expressed in terms of parameters evaluated  far upstream of the reconnection region.  

 \section*{Appendix C: Connection to results of resistive-MHD }

A critical reader may argue that the result in Eq.~\eq{Erec2} was just obtained two different ways, and the statement in Eq.~\eq{ErecErec} therefore becomes trivial. However, here it should be noted that Eq.~\eq{Erec2} was obtained based momentum balance constraints, while $\Delta p_{eyy,E}\simeq -e n l_u E\scr{rec}$ applied to the RHS of  Eq.~\eq{Erec2} represents the separate issues related to electron heating.  
As such, Eq.~\eq{ErecErec} can be considered  a meaningful mathematical expression describing the independence of $E\scr{rec}$ from the EDR dynamics, valid for any $E\scr{rec}$ sufficiently small that it does not alter the basic dynamics that lead to Eq.~\eq{Erec2}.

We also note that our  interpretation of Eqs.~\eq{Erec2} and \eq{ErecErec}
is not in conflict with results from resistive fluid models.
 For the particular case of an MHD plasma with resistivity $\eta$, the reconnection electric field is $E\scr{rec}=\eta J$, and the heating rate, ${\bf E} \cdot {\bf J}$, scales proportionally to $E\scr{rec}^2$. Power-balance arguments can then be invoked to quantify the much slower rate of reconnection characteristic of resistive MHD  \cite{Manheimer:1984,liu:2022}. In such fluid models with an imposed isotropic pressure, it is (by definition) not possible for $E\scr{rec}$ to change $p_{eyy}$ without introducing identical changes in 
$p_{exx}$  (and $p_{ezz}$). Therefore, Eq.~\eq{Erec2} is then
 not applicable and the analysis applied in resistive MHD has no implications on the collisionless anti-parallel case studied here.
In fact, as shown in Ref.~\cite{le:2016twostage}, when the energy continuity equation is  applied to the collisionless EDR of anti-parallel reconnection, we obtain a prediction of the  net electron heating level across the EDR, which does not impose a constraint on $E\scr{rec}$.

For the related case of collisionless guide-field reconnection, the structure of the EDR is qualitatively different as it does not include the jets of meandering electrons \cite{le:2009}. Further investigations are therefore needed to determine if the present framework can be generalized to scenarios including an out-of-plane guide magnetic field.

\section*{Appendix D: Characteristic length scales of the simulation domains}

\begin{table}[h]
\caption{\label{tab:table1}
Parameters describing the numerical simulation sizes in the $x$-direction. In the $z$-direction the domains all have half the size of the $x$-direction. For example, for $m_i/m_e=1836$ the domains are characterized by $L_x/d_{ep} \times L_z/d_{ep} = 1000\times 500$.} 
\begin{ruledtabular}
\begin{tabular}{ |c|c|c|c|c|c| } 
 \hline
 \multicolumn{6}{ |l| }{Table I.A: $L_x/d_{ep}$,   $L_x/d_{ip}$, \# of $x$-cells}  \\   \hline
 \multicolumn{2}{|r|} {$m_i/m_e=100$\,\,} & 200 & 400 & 800 & 1836   \\  
  \hline
 $L_x/d_{ep}$ &500 & 707 & 1000 & 1000 & 1000 \\ 
  \hline
 $L_x/d_{ip}$ & 50 & 50 & 50 & 35.35 & 23.34 \\ 
   \hline
 \# of $x$-cells & 3960  & 3960 & 5632 & 5632 & 5632 \\ 
 \hline
 \multicolumn{6}{ c }{  }  \\   \hline
 \multicolumn{6}{ |l| }{Table I.B: $L_x/d_{e\infty}$}  \\   \hline
   \multicolumn{2}{|r|} {$m_i/m_e=100$\,\,} & 200 & 400 & 800 & 1836   \\   \hline
 $\beta_{e\infty}=2^{-1}$ & 433 &  612 &  866 &  866 &  866  \\  \hline
  $2^{-2}$ & 387 &  548 &  775 &  775 &  775 \\   \hline
  $2^{-3}$ & 327 &  463 &  655 &  655 &  655 \\ \hline
  $2^{-4}$ & 261 &  369 &  522 &  522 &  522\\ \hline
  $2^{-5}$ & 199 &  281 &  397 &  397 &  397\\  \hline
  $2^{-6}$ & 146 &  207 &  293 &  293 &  293\\  \hline
  $2^{-7}$ & 106 &  150 &  212 &  212 &  212 \\    \hline
  \multicolumn{6}{ c }{  }  \\   \hline
 \multicolumn{6}{ |l| }{Table I.C: $L_x/d_{i\infty}$}  \\   \hline
 \multicolumn{2}{|r|} {$m_i/m_e=100$\,\,} & 200 & 400 & 800 & 1836   \\   \hline
 $\beta_{e\infty}=2^{-1}$ &  43.3 &  43.3 &  43.3 &  30.6  & 20.2  \\  \hline
  $2^{-2}$ &  38.7 &  38.7 &  38.7  & 27.4 &  18.1 \\   \hline
  $2^{-3}$ & 32.7  & 32.7  & 32.7 &  23.1 &  15.3 \\   \hline
  $2^{-4}$ & 26.1  & 26.1  & 26.1 &  18.5 &  12.2 \\   \hline
  $2^{-5}$ & 19.9  & 19.9  & 19.9 &  14.0 &   9.3 \\   \hline
  $2^{-6}$ & 14.6  & 14.6  & 14.6 &  10.4 &   6.8\\   \hline
  $2^{-7}$ & 10.6  & 10.6  & 10.6 &   7.5  &  4.9 \\  \hline
\end{tabular}
\end{ruledtabular}
\end{table}

\clearpage

\bibliography{referencesTail}

\end{document}